\documentclass[showpacs,preprintnumbers,amsmath,reprint,aps,prx,superscriptaddress]{revtex4-2}

\usepackage{bm}
\usepackage{graphicx}
\usepackage{color}
\usepackage{amsmath}
\usepackage{hyperref}
\usepackage{epstopdf}
\usepackage{upgreek}
\usepackage{mathptmx, textcomp}
\usepackage[latin1]{inputenc}
\usepackage[T1]{fontenc}
\usepackage{array, multirow}
\usepackage[table]{xcolor}
\usepackage{booktabs}
\usepackage{longtable}

\usepackage[normalem]{ulem}
\newcommand{\vib}{\text{\usefont{OML}{lmr}{m}{it}\symbol{118}}}

\newcommand{\commentOut}[1]{}

\begin{document}

\title{Energy-scaling of the product state distribution for three-body recombination of ultracold atoms}

\author{Shinsuke Haze} \email{shinsuke.haze.qiqb@osaka-u.ac.jp}
\affiliation{Institut f\"{u}r Quantenmaterie and Center for Integrated Quantum Science and
	Technology IQ$^{ST}$, Universit\"{a}t Ulm, 89069 Ulm, Germany}
\author{Jos\'{e} P. D'Incao}
\affiliation{Institut f\"{u}r Quantenmaterie and Center for Integrated Quantum Science and
	Technology IQ$^{ST}$, Universit\"{a}t Ulm, 89069 Ulm, Germany}
\affiliation{JILA, NIST, and the Department of Physics,
University of Colorado, Boulder, CO 80309, USA}	
	\author{Dominik Dorer}	
\affiliation{Institut f\"{u}r Quantenmaterie and Center for Integrated Quantum Science and
	Technology IQ$^{ST}$, Universit\"{a}t Ulm, 89069 Ulm, Germany}
\author{Jinglun Li}	
\affiliation{Institut f\"{u}r Quantenmaterie and Center for Integrated Quantum Science and
	Technology IQ$^{ST}$, Universit\"{a}t Ulm, 89069 Ulm, Germany}
		\author{Markus Dei\ss}
\affiliation{Institut f\"{u}r Quantenmaterie and Center for Integrated Quantum Science and
	Technology IQ$^{ST}$, Universit\"{a}t Ulm, 89069 Ulm, Germany}
\author{Eberhard Tiemann}
\affiliation{Institut f\"ur Quantenoptik, Leibniz
Universit\"at Hannover, 30167 Hannover, Germany}
\author{Paul S. Julienne}
\affiliation{Joint Quantum Institute, University of Maryland, and the National
Institute of Standards and Technology (NIST), College Park, MD 20742, USA}
		\author{Johannes Hecker Denschlag} \email{johannes.denschlag@uni-ulm.de}
\affiliation{Institut f\"{u}r Quantenmaterie and Center for Integrated Quantum Science and
	Technology IQ$^{ST}$, Universit\"{a}t Ulm, 89069 Ulm, Germany}

\date{\today}

\begin{abstract}	
Three-body recombination is a chemical reaction where the collision of three atoms leads to the formation of a diatomic molecule. In the ultracold 
regime it is expected  
that the production rate of a molecule generally decreases with its binding energy $E_b$, however, its precise dependence and the physics governing it have been left unclear so far.   
Here, we present a comprehensive experimental and theoretical study of the energy dependency for three-body recombination of ultracold Rb.
For this, we determine production rates for molecules in a state-to-state resolved manner, with the binding energies $E_b$ ranging from 0.02 to 77 GHz$\times h$.
We find that the formation rate approximately scales as $E_b^{-\alpha}$, where $\alpha$ is in the vicinity of 1.
The formation rate typically varies  only within a factor of two  for different rotational angular momenta of the molecular product,
apart from a possible centrifugal barrier suppression for low binding energies. 
  In addition to numerical three-body calculations we present a perturbative model which reveals the physical origin of the energy scaling of the formation rate. 
  Furthermore, we show that the scaling law potentially holds universally for a broad range of interaction potentials.
\end{abstract}

\maketitle
\section{Introduction}
\label{sec:intro}

When a molecule is formed in a chemical reaction there are often thousands of quantum states it
 can end up in, due to the various electronic, vibrational, rotational, and spin degrees of freedom.
Generally, the product population is not uniformly distributed over these possible product states, but
rather follows characteristic propensity rules. Finding and identifying propensity rules can provide deep insights on the basic principles which drive and govern specific reactions. 
Furthermore, the propensity rules can be used to develop predictions and approximations, especially when
full detailed calculations are highly complex, as, e.g., for reactions involving more than two atoms.

Propensity rules can be extracted experimentally from state-to-state measurements where the reactants are prepared in well defined quantum states and product states are detected in a quantum state resolved way.
In recent years, there has been rapid progress in the methodology of state-to-state chemistry using atomic and molecular beams \cite{Yang2007, Pan2017, Jankunas2015, Meerakker2012} or ultracold samples \cite{Liu2020, Haerter2013b}. Individual partial waves of product states have been resolved (see, e.g., \cite{Paliwal2021, deJongh2020, Beyer2018}) and spin conservation propensity rules have been observed with hyperfine and rotational states \cite{Haze2022, Wolf2017, Wolf2019, Hu2021, Liu2021}.

Three-body recombination is one of the 
most fundamental and ubiquitous chemical reactions. In a collision of three atoms, two combine to form a molecule and the third atom enables the dissipation of the released energy. The released
energy consists of the initial collision energy plus the molecular binding energy $E_b$ and is converted into relative motion between the molecule and the third atom. 
Experiments have shown that three-body recombination at ultracold temperatures  generally produces the
most weakly-bound molecular state (see, e.g. \cite{Weber2003, Jochim2003, Wolf2017, WangJChemPhys}).
 Semi-classical and fully quantum mechanical treatments have been carried out. They generally indicate that there is a propensity towards weakly-bound 
 molecular product states \cite{wang2011PRA, Perez2014, Wolf2017, Yuen2020}.  However, precisely how the molecular production rate decreases with the binding energy has not been clarified yet. 
 Reference \cite{Nesbitt2012}, e.g., suggested the suppression to be exponential in triatomic reactions.
 A recent calculation of three-body recombination of 
hydrogen atoms at room temperature predicted a molecular production rate $\propto E_b^{-1.5} $  
for recombination towards deeply-bound molecules, which was enhanced by the Jahn-Teller effect \cite{Yuen2020}.

Here, we investigate the rate decrease both experimentally as well as theoretically 
by studying three-body recombination of $^{87}$Rb atoms at ultralow collision energies.
We find a  $E_b^{-\alpha}$ power law  for the
molecular production rate where the exponent $\alpha$ is close to 1. 
This result differs from a previous scaling estimate of $\alpha = 1/2$ which was based on studying bound states in a limited range of binding energies \cite{Wolf2017}. We have now extended this range by roughly a
factor of ten, both on the experimental and theoretical side.
 Our experimental data comprise thirty final quantum channels of detected molecules with binding energies of up to $E_b = 77\:\textrm{GHz} \times h$.

Our numerical  calculations for the  three-body recombination rates $L_3 (E_b)$ are in remarkable  agreement with our measurements, especially for those product channels where molecules with low rotational angular momentum $L_R$ are formed.
Besides the general $E_b^{-\alpha}$ trend of $L_3 (E_b)$ the calculations also reproduce prominent deviations from this trend at particular binding energies $E_b$. These deviations might be interpreted as interference effects of various kinds.
 
\begin{figure*}[t]
	\includegraphics[width=2\columnwidth]{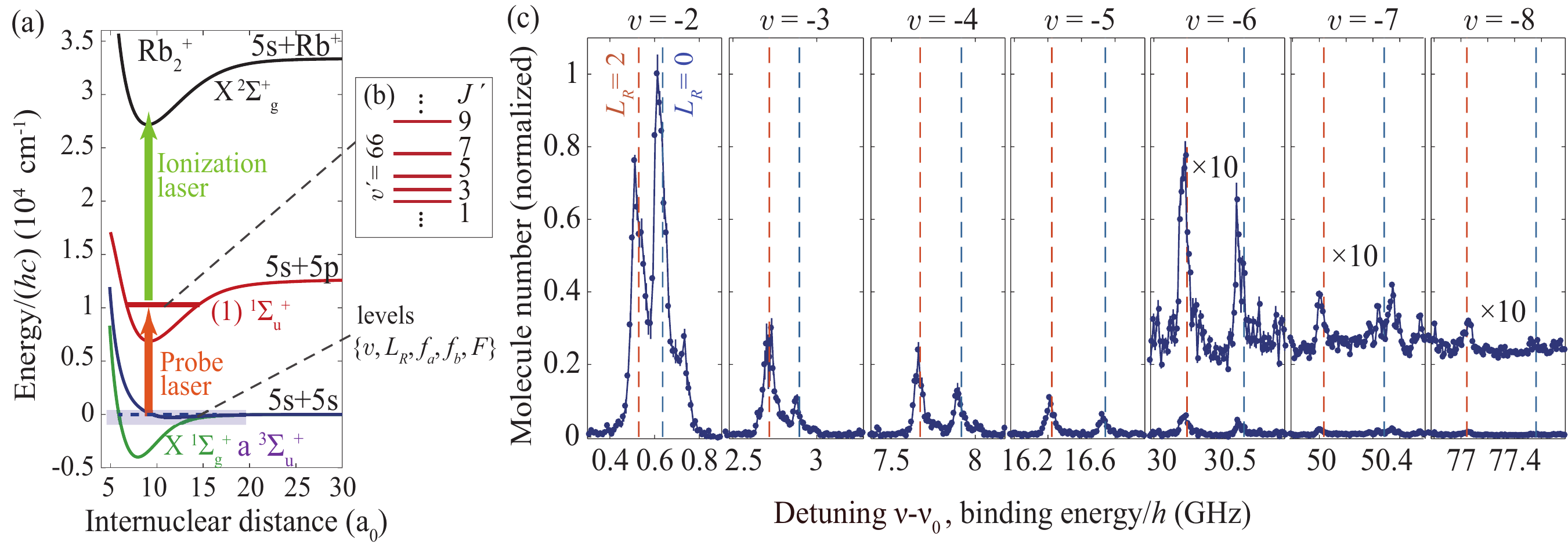}
	\caption{
	(a) Two-color REMPI scheme for state-selective detection of molecules.  Probe  and ionization lasers have a wavelength of $1065\:\textrm{nm}$ and $544\:\textrm{nm}$, respectively. (b) Rotational level structure of the relevant  molecular states with negative parity. (c) Various segments of REMPI detection signals of ($\vib, L_R=0,2$) product molecules as a function of the probe laser frequency $\nu$. Here, $\nu_0=281445.045\:\textrm{GHz}$, corresponding to the transition from the 5s 5s asymptote to $\vib' = 66, J' = 1$. The vibrational quantum numbers $\vib$ are given on the top of the figure and the rotational quantum numbers $L_R$ are indicated by the color coding of the dashed vertical lines ($L_R=0$ and $2$ in blue and red, respectively). These dashed lines are expected frequency positions for the given states obtained from coupled-channel calculations. For smaller signals ($\vib = -6,-7, -8$) magnifications by a factor of 10 are also shown. The data for the most-weakly bound state $\vib=-1,L_R=0$ are not presented, because the corresponding line is largely drowned by a neighboring photoassociation line \cite{Wolf2017}. }
	\label{fig2}
\end{figure*}

 Our perturbative model indicates that for each angular momentum $L_R > 0 $ there is a critical binding energy
$E_c (L_R)$ so that for $ E_b  > E_c (L_R)$
the trend of the partial recombination rate 
will be described by $L_3(E_b,L_R) = c  E_b^{-\alpha}$,
where $c$ is a constant. We find that the factor $c$ is roughly independent of $L_R$.
For 
$ E_b   <  E_c (L_R)$ there is a suppression of $L_3(E_b,L_R)$ which can be explained 
as the effect of an angular momentum barrier in the exit channel. 
As a result, this suggests that  only molecular states with small $L_R$ will significantly contribute to molecular production at low binding energies $E_b$. 

Finally, we show that the $E_b^{-\alpha}$ scaling law can be also derived theoretically in an analytic, perturbative approach. We find that within this approach the $E_b^{-\alpha}$ scaling is quite independent on the long-range behavior of the interaction potential between two atoms.  Specifically, potentials with a power law 
 tail $-C_n/r^n$ for $n=3,4$ or 6, or the Morse potential, as well as the contact potential have $\alpha$-values in the range $[0.91, 1]$. 
 
Within the framework of the perturbative calculations 
the scaling of the rate constant $L_3(E_b)$ is largely determined by $|\phi_d(\sqrt{E_b\,m/3})|^2$, 
where $\phi_d$ is the diatomic molecular wave function in momentum representation and $m$ is the atomic mass. It turns out that this part of the momentum wave function is linked to the molecular wave function in real space in the vicinity of the classical outer turning point of the molecular potential (at energy $-E_b$).
This indicates that the outer classical turning point marks
a typical distance for the recombination to occur.

\section{Experiment}
\label{sec:experiment}

\begin{figure}[t]
	\includegraphics[width=0.8\columnwidth]{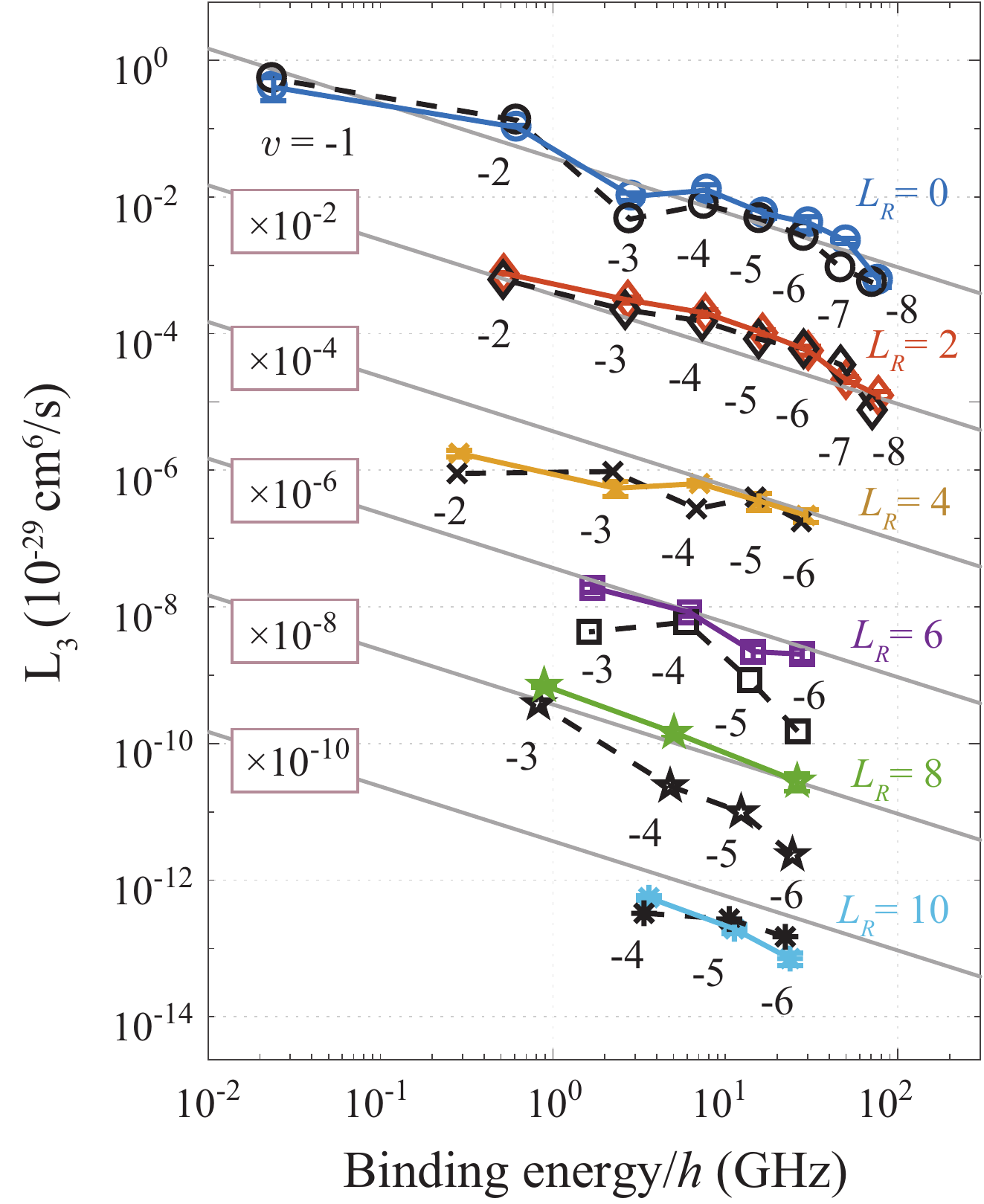}
	\caption{Measured (scaled) and calculated rate constants $L_3(\vib,L_R)$ for product molecules $(\vib,L_R)$ as a function of their binding energy $E_b$. The rotational quantum numbers $L_R$ are indicated by different plot symbols (black: calculations; colors: experiment). For each sequence of states with the same $L_R$ the vibrational quantum number $\vib$ is given below the data points. 
 For better visibility the data for each $L_R$ are shifted in vertical direction by multiplying them with $10^{-L_R}$. The gray solid lines show the energy scaling $(E_b/\textrm{GHz}\times h)^{-0.8} \times 0.2\times 10^{-29}\:\textrm{cm}^6/\textrm{s}$.  
	}
	\label{fig3}
\end{figure}

In our experiments we prepare an ultracold cloud of $5\times10^6$  $^{87}$Rb atoms in a far-detuned 1D optical lattice trap ($\lambda = 1064\,$nm, trap depth $ \approx 10\,\mu$K$\times k_B$) combined with an optical dipole trap so that we obtain a trap frequency of $2\pi \times 23 $Hz in the transverse direction. The atoms are spin-polarized in the hyperfine state $f=1,m_f=-1$ of the electronic ground state and have a temperature of about $750\:\text{nK}$. Our measurements are carried out at a low external magnetic field of about $4\:\textrm{G}$. We hold the atom cloud in the trap for a duration of 500$\:$ms during which Rb$_2$ molecules are spontaneously produced via three-body recombination in the coupled  $X^1\Sigma_g^+-a^3\Sigma_u^+$ molecular complex, below the $5S_{1/2}+5S_{1/2}$ atomic asymptote. The molecules are state-selectively ionized via resonance-enhanced multiphoton ionization [REMPI] (see Fig.$\:$\ref{fig2}(a) and Appendix \ref{appendix:REMPI} for details), and then trapped and detected as ions in a Paul trap at a distance of 50 $\mu$m (see Appendix \ref{appendix:counting} for details). In brief, a first REMPI laser (the probe laser) resonantly excites such a  molecule to the intermediate level $\vib ' =66, \, J'$  of the state $A^1\Sigma_u^+$ using a wavelength of about 1065$\:$nm \cite{Deiss2015,Drozdova2013}. Here, $\vib ' $  is the vibrational quantum number and  $J'$ is the total angular momentum quantum number excluding nuclear spin. From the intermediate level a second laser (the ionization laser) at a wavelength of about 544$\:$nm resonantly excites the molecule to a state above the Rb$_2^+$ ionization threshold, so that the molecule can autoionize. In one experimental run we can detect and count up to $\approx$ 70 ions in the Paul trap. The ion number scales linearly with the molecule number. 
The corresponding scaling factor $\eta$ is the detection efficiency of a molecule. 
As discussed in Appendices \ref{appendix:REMPI} and \ref{appendix:Conversion}, $\eta $  is roughly constant over the range of bound states investigated in this work, and its value is  $\eta \approx 4.8\times10^{-3}$.
A REMPI spectrum of a particular product state is obtained by scanning the probe laser frequency in steps of typically 5$\:$MHz. 
Our setup features an improvement of the product state signals and the sensitivity by a factor of $\approx 25$ as compared to previous work \cite{Wolf2017}, extending our detection range of binding energies to about $80\:\textrm{GHz}\times h$, which was instrumental for the present work (for details, see Appendix \ref{appendix:boost}).

In the following we specify molecular states by their vibrational quantum number $\vib$ and their rotational quantum number $L_R$ only, which is sufficient due to the conservation of the hyperfine spin state in the reaction process \cite{Noteconservation}. Figure \ref{fig2}(c) shows product state spectra for molecules with $L_R = 0$ or $2$, and with $\vib$ ranging from $ -2$ to $-8$.
We note that whenever  $\vib$ is negative,
it is counted downwards from the atomic $f_a = f_b = 1$, $m_{fa} =m_{fb} = -1$ asymptote, starting with $\vib = -1$ for the most weakly bound vibrational level.  
The most deeply bound state, ($\vib = -8, L_R=0$), has a binding energy of $77\:\textrm{GHz} \times h$. Here, all signals are obtained using the same intermediate state $J'=1$ for REMPI. The frequency reference $\nu_0$ corresponds to the photoassociation transition towards this intermediate state such that, at a resonance position, $(\nu-\nu_0)\times h$ directly represents the binding energy of the initially produced molecular state. Our data clearly show that the production rate of molecules for a given rotational level $L_R$ generally drops   with the binding energy $E_b$. The drop is significant over the investigated range of $E_b$.
The relative strength of $L_R = 0$ and $2$ signals, however, can vary for different 
vibrational levels $\vib$. 
 Large molecular signals as, e.g., obtained for $\vib=-2$ correspond to 63(3) produced ions per run whereas typical background signals are 0.69(0.15) ions per run (see also Fig.$\:$\ref{fig:App6} in Appendix \ref{appendix:boost}).  In the measurements of Fig.$\:$\ref{fig2}(c) the number of repetitions of the experiment per data point was gradually increased from 5 to 40 for increasing binding energy, in order to improve the visibility of  smaller signals. We assign the signals in our REMPI spectra by comparing their frequency positions to those obtained from
 close-coupled channel calculations (see, e.g., \cite{Wolf2017,Haze2022}) and by observing
 characteristic rotational ladders, since any molecular state with $L_R>0$ can be detected via  two different rotational states $J'=|L_R\pm 1|$ [see Appendix \ref{appendix:ConsistencyCheck}, not shown in Fig.$\:$\ref{fig2}(c)]. 
 The deviations between calculated [dashed vertical lines in Fig.$\:$\ref{fig2}(c)] and measured resonance frequency positions are typically smaller than $\sim 30\:\textrm{MHz}$ and arise mainly from daily drifts of our wavelength meter.
 
 \section{Quantitative analysis}
\label{sec:numerical model}

We now carry out a quantitative analysis of the observed population distribution. Figure \ref{fig3} shows measured (scaled) and calculated partial rate constants $L_3(E_b) \equiv L_3(\vib,L_R)$ for the production of $(\vib,\, L_R)$ molecules at a temperature of 0.8$\:\mu$K. The experimental values
for $L_3(\vib,L_R)$ were obtained 
by multiplying the measured 
ion numbers with a single calibration factor for all detected ($\vib,L_R$) states. This factor  was chosen to 
optimize the agreement between experiment and theory (Appendix \ref{appendix:Conversion}).
 As can be seen from Fig.$\:$\ref{fig3}, the theoretical predictions reproduce remarkably well the relative strengths of the state dependent rates $L_3(\vib, L_R)$ obtained from the experiments. The calculations are based on solving the three-body Schr\"{o}dinger equation in an adiabatic hyperspherical representation \cite{dincao2018JPB,wang2011PRA}, using a single-spin model as described in Appendix \ref{appendix:TBmodelRb87} (see also Ref.$\:$\cite{Wolf2017}).
Within our model the $^{87}$Rb atoms interact via pairwise additive long-range van der Waals potentials with a scattering length of 100.36 a$_0$ \cite{Strauss2010}. The potentials are truncated  and support $15$  $L_R=0$ molecular bound states, and a total of 240 bound states. We calculated the theory data point for $L_{R}=8$, $\vib=-3$ using a model potential with 12 $s$-wave bound states because for 15 $s$-wave bound states a numerical instability occurs specifically for this level.  
 
 Figure$\:$\ref{fig3} reveals that the $L_3(\vib,L_R)$  rate roughly follows  the overall scaling of  $E_b^{-\alpha}$ for all rotational states. A fit analysis to the experimental data yields a scaling factor $\alpha={0.80(\pm 0.14)}$ (see gray solid lines), while the fit to the theoretical data yields $\alpha={0.77(\pm 0.10)}$. We point out that all gray solid lines 
 in Fig.$\:$\ref{fig3} correspond to exactly the same function, 
 $(E_b/\textrm{GHz}\times h)^{-0.8} \times 0.2\times 10^{-29}\:\textrm{cm}^6/\textrm{s}$. For better visibility these lines along with the respective data points have been shifted in vertical direction by multiplying them
 with $10^{-L_R}$. We notice that the measured data for $L_R = 2$ are all located above the gray line while
 the data for $L_R = 4$  or $10$ are all below the gray line. This indicates that there is a systematic dependence of $L_3(E_B)$ on $L_R$, as already discussed in 
 \cite{Wolf2017}. Nevertheless, considering the overall range of $L_3$ and $L_R$ in our data, this variation of 
 $L_3(E_B)$ with $L_R$ is still comparatively small, typically within a factor of 2.  Therefore, to a first approximation, the production rate does seem
 to be quite independent of the molecular rotation $L_R$. This fact might be somewhat counterintuitive given that the atoms initially collide with vanishing angular momenta and therefore products with small angular momenta would seem to be naturally preferable.  
 
We note that there are considerable variations around the general $E_b^{-\alpha}$ scaling trend. For example, the rate for the state $\vib=-3, L_R=0$ is significantly lower than the rate for the more-deeply bound state $\vib=-4,L_R=0$.  Remarkably,  even such individual variations are largely reproduced by our numerical calculations. 
In general, the theoretical and experimental data curves are very similar, especially for the low rotational states $L_R=0$ and $L_R=2$. This suggests that our three-body model is quite accurate
and that it should in principle be capable to track down how the deviations from the general scaling come about in individual cases. 
For example, we point to the experimental and theoretical data points for $\vib = -2$ and $L_R=4$, which are located
below the $E_b^{-\alpha}$ scaling trend. This suppression may be due to an angular momentum barrier effect which we will discuss
in Section \ref{sec:hiPartialWave}.

\section{Perturbative approach}
\label{sec:perturbative}

\begin{figure}[t]
 	\centering
 	\resizebox{0.53\textwidth}{!}{\includegraphics{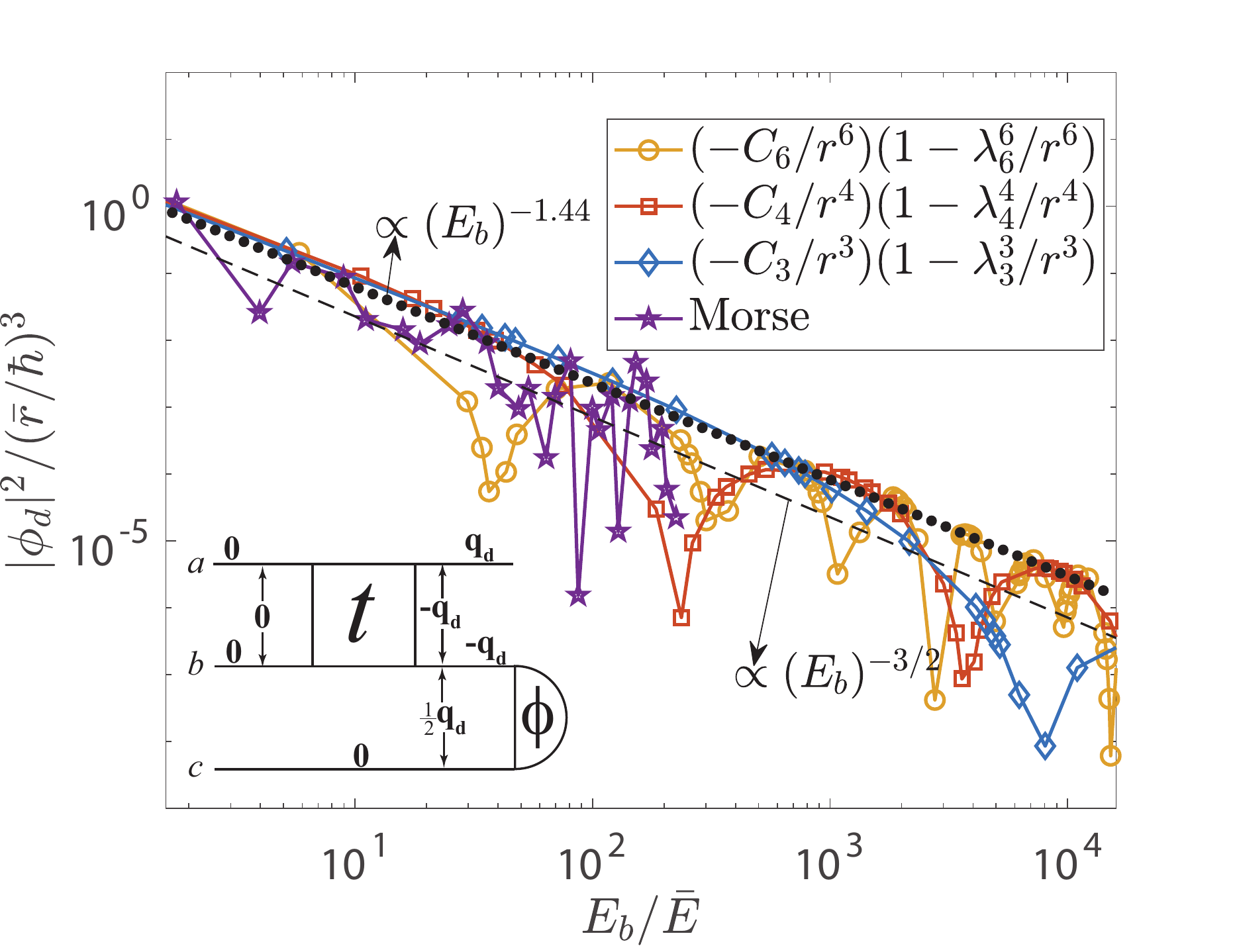} }
 	\caption{
 	Plot of $|\phi_d(\sqrt{E_b\,m/3})|^2 \propto L_3(E_b)/\sqrt{E_b}$ versus $E_b$ for various two-body potentials.  These are the potentials with the power-law tail, $\sim -C_n/r^n$ for $n=3,4$ or 6, and the Morse potential $D_e[e^{-2a(r-r_e)}-2e^{-a(r-r_e)}]$ (see legend).  We use $\bar{E}=\hbar/m \bar{r}^2$, with $\bar{r}=\frac{1}{2}(m C_n/\hbar^2)^{1/(n-2)}$ for the power-law potentials and $\bar{r}=a$ for the Morse potential, respectively. The values for $\lambda_n$ and $D_e$ are chosen such that the potentials have 14 or 15 $s$-wave bound states. The shown 
 	points correspond to bound states with $\vib \geq -13$.
 	In order to map out the function $|\phi_d(\sqrt{E_b\,m/3})|^2$ better  we show data points for four different values for  $\lambda_n$ (for each $n$) or for $D_e$ (see text). The dotted line is a power-law fit to the upper envelope of the data points for the $-C_6/r^6$ potential as well as the other potentials. 
   The dashed line represents the scaling for contact interactions. For $n = 6$ the energy range considered in this figure corresponds to [0.01 \dots 100] GHz$\times h$ for $^{87}$Rb atoms. The inset describes the scattering process in the perturbative approximation (see Appendix \ref{appendix:AGS}).  Horizontal lines represent atoms and the numbers above these lines denote single atom momenta. The relative momentum between two atoms is indicated by a number that connects to the corresponding horizontal lines by arrows.
 	}\label{fig:phi_decay}
 \end{figure}

In order to gain a deeper insight into the observed energy scaling we discuss in the following
a perturbative  model for the partial three-body recombination rates $L_3(\vib, L_R)$.
Generally, the rates $L_3(\vib, L_R)$ towards each specific molecular product $d = (\vib, L_R)$ are given by \cite{Li:2022} (see also Appendix \ref{appendix:AGS}) 
\begin{equation} \label{eq:L3a}
L_3(\vib, L_R)= \frac{12 \pi m}{ \hbar} (2 \pi \hbar)^6 q_{d} |\langle\psi_{f} | U_{0} (E) | \psi_{\rm{in}} \rangle |^2,
\end{equation}
where $m$ represents the mass of an atom. $| \psi_{\rm{in}}\rangle$ is the initial state, consisting of three free atoms each propagating as a plane wave with essentially vanishing momentum. 
$|\psi_{f}\rangle$ is the final state of a free atom and a free molecule. Atom and molecule are asymptotically propagating as  plane waves with  relative
 momentum $q_{d}$ which is fixed by the molecular binding energy $E_b$ and the total energy $E$ of the three-body system 
 via $\frac{3q_{d}^2}{4m}-E_b=E$, where we use the  center of mass system as a reference. 
 In Eq.$\:$(\ref{eq:L3a}), $ U_{0} (E)$ represents a  three-body transition operator which describes the transition process between the states. It can be approximated 
 by a perturbative expansion (Appendix \ref{appendix:AGS}) derived from the Alt-Grassberger-Sandhas (AGS) equation  \cite{Alt:1967,Secker:2021,Li:2022}. 
 To the leading order of the expansion, we have 
 a process where atoms $(a,b)$  of the three free atoms $(a,b,c)$ collide
 to exchange a momentum $\mathbf{q}_d$. During this collision atom $(b)$ is scattered into a molecular bound state with atom $(c)$. 
This is shown schematically in the inset of  Fig.$\:$\ref{fig:phi_decay}. The initial momenta of the atoms are 0.
 After the collision atom $(a)$ remains free and carries away the momentum $\mathbf{q}_d$ and a corresponding part of the released binding energy.
 The formed molecular bound state $\phi$  has a total momentum  $-\mathbf{q}_d$ and the relative momentum
 between its atomic constituents is  $(-\mathbf{q}_d/2)$. 
 Apart from constants, the result of the calculation is
  \begin{equation} \label{eq:L3b}
 L _3(\vib, L_R) \propto  q_d \, \,  \, \left| \phi_d \left(\frac{1}{2}q_d \right) \right|^2 \, \, \,   |t_{\rm{h}}(q_d)|^2.
 \end{equation}
 Here, $\phi_d(p=\frac{1}{2}q_d)$ corresponds to the radial part of the molecular wave function in momentum space. It is normalized according to  $\int |\phi_d(p)|^2 p^2 dp=1$. The factor $t_{\rm{h}}(p'=q_d) \equiv$  $\langle p'= q_d|t^{s}(0)|p=0\rangle$ 
 is the matrix element of the $s$-wave component $t^s$ of the two-body transition operator $t$ for the two-body collision
 and we have set $E = 0$. 
 Here, $p$ ($p'$) represent the relative momenta of the incoming (outgoing) two colliding atoms, respectively. 
 Within the perturbative approximation, the $E_b$ scaling of $L_3(\vib, L_R)$ can only result from two-body quantities, i.e., $E_b$, $\phi_d$ and $t_{\rm{h}}$. Since $E \approx 0$, one obtains $q_d \approx 2 \sqrt{E_b \, m / 3 }$.
 
  In order to analyze the scaling of $L_3(\vib, L_R)$ with the molecular binding energy $E_b$,
 we discuss $\phi_d(\sqrt{E_b\,m/3})$ and $t_{\rm{h}}(2\sqrt{E_b\,m/3})$ separately. We find that $t_{\rm{h}}(p)$ oscillates but its amplitude varies only gently with $p$ 
  until the deeply-bound states are reached (see 
  Fig.$\:$\ref{fig:th}(a) in
  Appendix \ref{appendix:AGS}). Therefore, $t_{\rm{h}}(2\sqrt{E_b\,m/3})$ cannot strongly contribute to an overall scaling with $E_b$ for the three-body recombination rate. 
 In contrast to that,  $\phi_d(\sqrt{E_b\,m/3})$ which is obtained from two-body bound state calculations, vanishes quickly with increasing $E_b$.
 This is shown in Fig.$\:$\ref{fig:phi_decay} (yellow data points) for atoms interacting via the van der Waals potential. 
Besides the overall decrease of  $|\phi_d(\sqrt{E_b\,m/3})|^2$ for growing $E_b$, there are also oscillations. The sharp drops in these oscillations correspond to the nodes of the various momentum wave functions $\phi_d$ for the bound states with energy $E_b$. 
While the oscillations lead to some scatter of the data, the upper envelope  of the data points indicates an overall power law scaling of the 
amplitude. 
 A fit to this envelope (dotted line in Fig.$\:$\ref{fig:phi_decay}) gives $|\phi_d(\sqrt{E_b\,m/3})|^2\propto E_b^{-1.44\pm 0.03}$  which yields $L_3\propto E_b^{-0.94 \pm 0.03}$. This result agrees quite well with our full calculations from Fig.$\:$\ref{fig3}.
We note that in the shown energy range there are only 13 bound states in the van der Waals potential, resulting in 13 data points. 
In order to map out in more detail the functional form of  $|\phi_d(\sqrt{E_b\,m/3})|^2$ in Fig.$\:$\ref{fig:phi_decay}
 we have slightly varied $\lambda_6$ over four different values (while keeping the number of bound states in the potential constant).
 The variation in $\lambda_6$ leads to variations of $E_b$ and therefore also of $\phi_d$ and the scattering length $a$. 
 When we present all these data points together, a quasi-continuous curve is obtained. We note that the oscillating amplitude of $t_{\rm{h}}(p)$, which is nearly constant for a fixed scattering length  $a$, can depend on $a$.
 
 \section{ Energy scaling for general long-range potentials}
\label{sec:analytical derivation}
Remarkably, we find that the scaling law is similar for a range of different two-body interaction potentials, such as
the Morse potential and potentials of the 
form $V(r) =  -C_{n}/r^n (1- \lambda_{n}^{n}/r^{n})$. Here, the parameter $n$ is typically  $n = 3, 4$ or  6, and
$\lambda_n$ is a short-range parameter which defines the inner barrier. The case $n = 6$ corresponds to the Lennard-Jones potential which was already discussed in the previous section. The corresponding functions $|\phi_d(\sqrt{E_b\,m/3})|^2$ are shown 
in Fig.$\:$\ref{fig:phi_decay}. Clearly, their envelopes roughly decrease in a similar manner, i.e. 
$  \approx E_b^{-1.44\pm 0.03}$. 
Furthermore, 
we also consider contact interactions between the atoms. For these, we use 
 $\phi_d(p)=\frac{2}{\sqrt{\pi}}\frac{(m \, E_b)^{1/4}}{p^2+m \, E_b}$ and  analytically obtain from Eq.$\:$(\ref{eq:L3b}) the scaling to be exactly
 $|\phi_d(\sqrt{E_b\,m/3})|^2\propto1/E_b^{3/2}$ (see black dashed line in Fig. \ref{fig:phi_decay}), corresponding to $L_3(E_b)\propto 1/E_b$. 
 Therefore, even the results for contact interaction are in relatively good agreement with our other
  numerical and the experimental results.
  
   \begin{figure}[t]
 	\centering
 	\resizebox{0.53\textwidth}{!}{\includegraphics{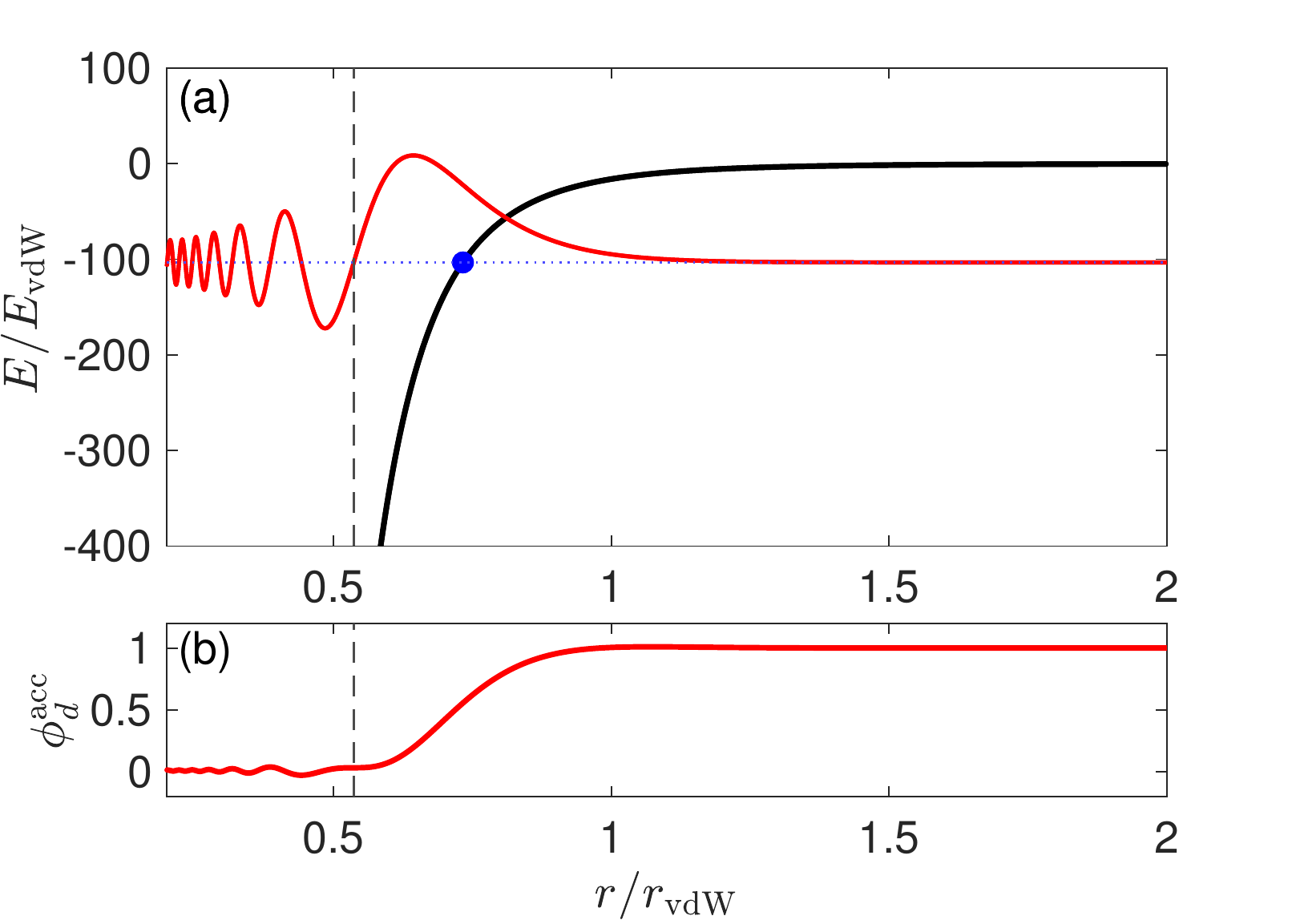}}
 	\caption{(a) Typical example for a molecular wave function (red solid line)
 	located in the $-C_6/r^6$ van der Waals  potential (black solid line). The blue horizontal dotted line and the blue circle indicate the molecular level position $-E_b$ and the corresponding classical outer turning point, respectively. The vertical dashed line shows the starting position of the last lobe of the molecular wave function. (b) shows the accumulated Fourier integral  $\phi^{\rm{acc}}_d(\sqrt{mE_b/3};r)$, see text. 
  $ E_\mathrm{vdW}  =\hbar/m r_\mathrm{vdW}^2$ is the van der Waals energy and $r_\mathrm{vdW} =\frac{1}{2}(m C_6/\hbar^2)^{1/4}$ is the van der Waals length.}
 \label{fig:lastlobe}
 \end{figure} 
  
In the following we discuss how this similar scaling for the different long-range potentials can be explained. We make use of approximate analytical wave functions for the molecular bound state. 
Let $\psi(r) = u(r)/r $ be the radial part of the molecular wave function with rotational angular momentum $L_R$. Here, $r$ is the internuclear distance between the two atoms. A typical example of the reduced radial wave function  $u(r)$ is shown in Fig. \ref{fig:lastlobe}(a).
The Fourier transform of $u(r)/r $ generates the molecular wave function in momentum space
\begin{eqnarray}
  \phi_d (p)&=&\sqrt{\frac{2}{\pi}}\int_0^{\infty} r j_{L_R}(pr/\hbar) {u}(r)dr ,
  \end{eqnarray}
   where  $j_{L_R}$ is the spherical Bessel function of the first kind of order $L_R$.
A numerical analysis shows 
 that the dominant contribution to $\phi_d(\sqrt{ E_b m/ 3})$ comes from the last lobe of 
 $u(r)$,
 which is located around the classical outer turning point $r_0$. 
In Fig. \ref{fig:lastlobe}(b) we show the Fourier integral
for the case $L_R = 0$
in an  
accumulated fashion 
$\phi^{\rm{acc}}(p;r)=\int_{0}^{r}\sin(pr/ \hbar)u(r)dr/\int_{0}^{\infty}\sin(pr/ \hbar )u(r)dr$,
 which verifies the
dominant contribution of the last lobe. 
 The turning point of the level [blue circle in Fig.
 \ref{fig:lastlobe}(a)] is determined by 
  $\tilde V(r_0) +E_b=0$, where  $ \tilde V(r) =  V(r) + \hbar^2 L_R (L_R + 1)/ (m r^2)$. The reduced radial wave function $u(r) = r \psi(r)$ in this region is approximated by 
 $\tilde u(r) =  \mathcal{N}^{1/2}\textrm{Ai} [s(r-r_0)]$, where Ai($x$) is the Airy function,  $s=(mD/\hbar^2)^{1/3}$,
  $D= d\tilde{V}(r)/ dr|_{r=r_0}$, and $\mathcal{N}$ is a normalization factor that ensures that $\tilde u(r)$ best matches
  $u(r)$ in the region. It turns out that to a good approximation 
  $\mathcal{N} = s N$, where $N$ is a constant independent of $E_b$ \cite{LastLobe}. 
We  Fourier transform $\tilde{u}(r)$ and obtain
  \begin{eqnarray}
  \tilde \phi_d (p)
 &=&\sqrt{\frac{2N\hbar^2}{\pi s p^2}}  g_d(p)\,,
  \end{eqnarray}
  where 
   \begin{equation}
  g_d(p)=\int_0^{\infty}\frac{p\tilde{r}}{s\hbar} j_{L_R}(p\tilde{r}/s\hbar)\textrm{Ai} (\tilde{r}-\tilde{r}_0)d\tilde{r}. 
  \end{equation}
  Here, $\tilde{r}=sr$, $\tilde{r}_0=sr_0$.  At $p=\sqrt{ E_b m/ 3}$, we get
  \begin{align}
  |\tilde \phi_d (\sqrt{ E_b m/ 3})|^2 &=\frac{6N\hbar^2}{\pi s mE_b}g_d^2(\sqrt{ E_b m/ 3}), 
  \label{eq:aphi}
  \end{align}
  which approximates $|\phi_d (\sqrt{ E_b m/ 3})|^2$. 
  \begin{figure}[t]
  	\centering
  	\resizebox{0.52\textwidth}{!}{\includegraphics{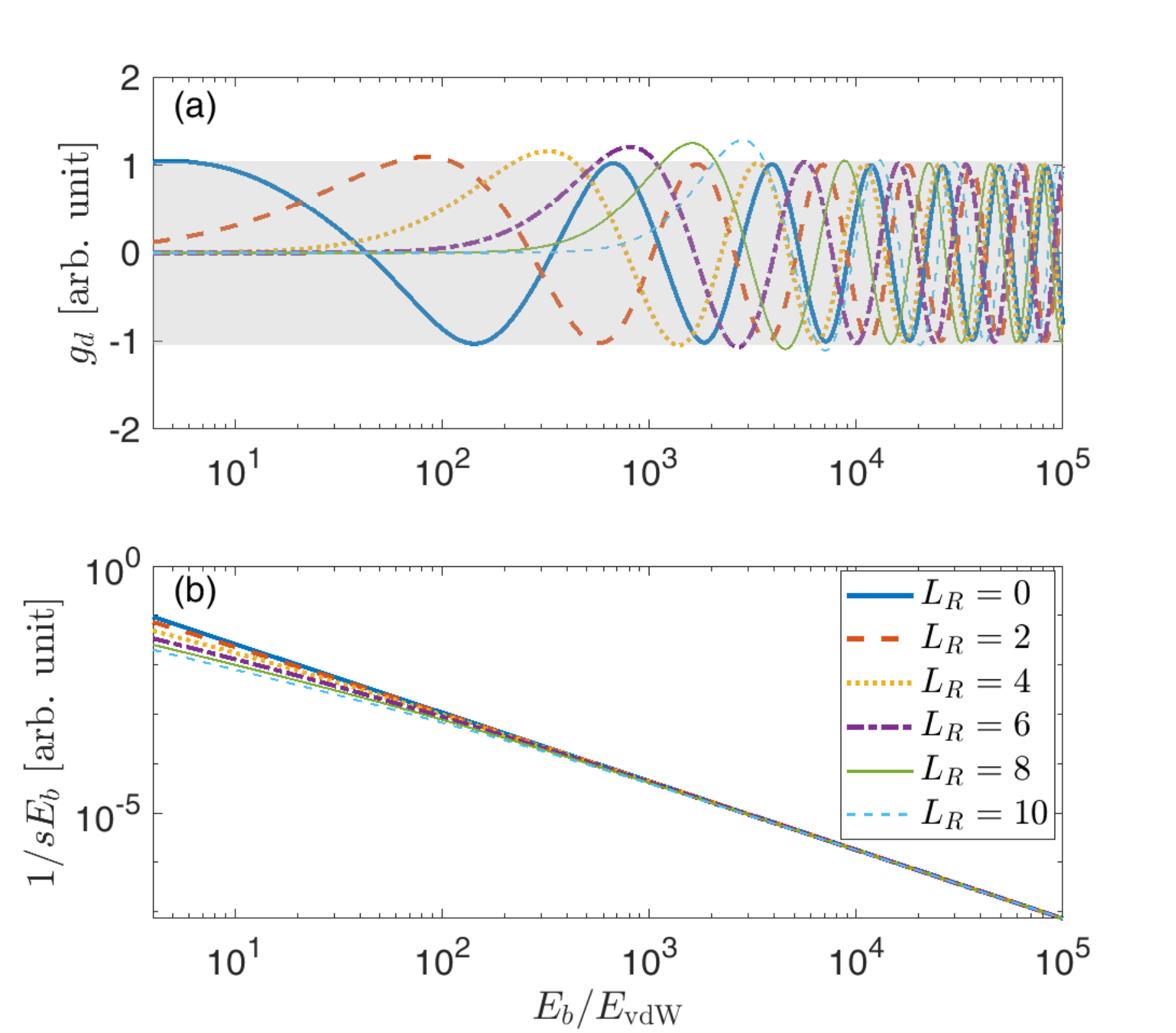} }
  	\caption{\label{fig:phai}
   (a) Plot of  $g_d(\sqrt{ E_b m/ 3}) $ as a function of $E_b$ for the partial waves $L_R \leq 10$ (see legend in (b)). The gray area highlights the range of the oscillation for $L_R=0$.
   (b) Plot of $1/(s E_b)$. 
   For both (a) and (b) we use the $-C_6/r^6$ potential.   }
  \end{figure} 
After this general discussion we now discuss the cases for $L_R = 0$ and
$L_R > 0$. 
 \subsection{Case: $L_R=0$}
For a $L_R=0$ molecular state and  $\tilde V(r)=-C_n/r^n$  the classical outer turning point is given by $r_0 = (C_n/ E_b)^{1/n}$. From the derivative of the 
potential $\tilde V (r)$ we obtain for the parameter $s$
(which we defined earlier in connection with the Airy function), 
\begin{align}
s &=(mn/\hbar^2)^{1/3}E_b^{(n+1)/3n}C_n^{-1/3n}  \label{eq:sLR0}.
\end{align}
Thus, the factor $6N\hbar^2/(\pi s mE_b)$  in Eq.~(\ref{eq:aphi})  scales as  $E_b^{-(4n+1)/3n}$.
 The other factor in Eq.~(\ref{eq:aphi}), $g_d(\sqrt{ E_b m/ 3})$, is plotted
in Fig.~\ref{fig:phai}(a) for $n = 6$, (blue line).  
   $g_d(\sqrt{ E_b m/ 3})$ oscillates between 1 and -1 with a constant 
 amplitude, as indicated by the gray area. Therefore, $g_d(\sqrt{ E_b m/ 3})$ does not contribute to an overall scaling of $|\phi_d (\sqrt{ E_b m/ 3})|^2$  with  $E_b$.
This is  similar as for $t_h(q_d)$ as mentioned in Sec. \ref{sec:perturbative}. 
 The variation of $g_d(\sqrt{ E_b m/ 3})$ merely leads to some scatter of $L_3$. Therefore, we  ignore $g_d(\sqrt{ E_b m/ 3})$ in the following discussion on scaling and 
  obtain,
  \begin{equation}
  |\phi_d (\sqrt{ E_b m/ 3})|^2 \propto E_b^{-(4n+1)/3n} =  E_b^{-\beta} . \label{eq:scale}
  \end{equation}
  For $n=3,4$ and 6, the exponent  $\beta = (4n+1)/3n$ takes the 
   values of 1.44, 1.42 and 1.39, which agree  very well with our numerical results in the perturbative approach. 
  For  the  scaling of $L_3$ we  have $L_3\propto E_b^{-\alpha}$, where $\alpha = \beta - 0.5$ \textcolor{magenta}.
   It is  remarkable that for any positive integer $n$ the exponents are constrained to a narrow range, i.e. $\beta \in$ [1.33, 1.67] and $\alpha \in$ [0.83, 1.17]. 
Considering that a real interaction potential can  typically  be expanded in terms of the $-C_n/r^n$ functions,
these ranges should be valid quite generally. 
 In fact, the range of $\alpha \in$ [0.83, 1.17] agrees with  the exponent $\alpha=0.8 \pm 0.14$ extracted from our experimental measurements
 within the range of uncertainty.

 \subsection{Case: $L_R>0$}
\label{sec:hiPartialWave}

We now consider the case of rotational angular momentum  $L_R>0$. Because for this case we do not obtain a simple analytical expression for 
$s$ as in Eq.~(\ref{eq:sLR0}), we only present numerical results. 
Figure \ref{fig:phai}(b) shows a plot of $1/(s E_b)$ for various  $L_R$ (and using the $-C_6/r^6$ potential as a
typical example).
 Clearly, 
all curves are quite similar, especially for large $E_b$. 
The functions  $g_d(\sqrt{ E_b m/ 3})$  are shown in Fig.~\ref{fig:phai}(a), as discussed before. They oscillate with a constant amplitude and
therefore do not contribute to the energy scaling for large energies $E_b$. 
As a consequence, the energy scaling is quite independent on the  rotational state $L_R$ of the molecule. Figure \ref{fig:phL} shows calculations for $|\phi_d(\sqrt{E_b\,m/3})|^2$ for various rotational angular momenta of the molecule. As a typical example we use  the Lennard-Jones potential. 
Similar as for  Fig.\ref{fig:phi_decay}  we have slightly varied $\lambda_6$ over four different values 
in order to increase the number of data points and to better map out $|\phi_d(\sqrt{E_b\,m/3})|^2$. 
The sudden drops in $|\phi_d(\sqrt{E_b\,m/3})|^2$ reflect nodes of the molecular wave function. 
For large enough binding energy $E_b$ all curves for the different values of $L_R$ follow the same power law $E_b^{-1.45}$ corresponding to an energy scaling of the partial rate constants (for fixed $L_R$) of  $L_3(E_b)\propto E_b^{-0.95}$. 

We note, however, that for $L_R>0$ and small enough binding energies, our calculations  in Fig.  \ref{fig:phL}  reveal a strong suppression of $|\phi_d|^2$ and therefore of $L_3(E_b)$.
This effect is due to the function  $g_d(\sqrt{ E_b m/ 3})$. 
 As shown in Fig. \ref{fig:phai} (a), for $L_R>0$,  $g_d(\sqrt{ E_b m/ 3})$ increases gradually with $E_b$ starting from 0.
 When it reaches its first maximum, it goes over to the previously discussed oscillatory behavior in the gray area, similar to the case of $L_R=0$. 
 As a consequence,  $|\phi_d (\sqrt{ E_b m/ 3})|^2$ is increasingly suppressed  for  $E_b \rightarrow 0 $, as observed in our numerical results. 
 This suppression can be understood as an effect of the angular momentum barrier. 
 In a simple picture, in order to create a molecule rotating with angular momentum $L_R$ at interparticle distance $r_0$ of the outer turning point,  a minimal momentum $p_c$
needs to be supplied of the order $p_c \approx \hbar L_R / r_0$. The minimal momentum $p_c$ translates into a minimal 
binding energy $E_c =3 p_c^2 / m \approx \hbar^2 L_R(L_R +1) / (m r_0^2)$. At the same time we have $E_c \approx C_6 / r_0^6-\hbar^2 L_R(L_R +1) / (m r_0^2)$. Combining these two equations to eliminate $r_0$ one can estimate
 the critical energy $E_c$ 
to be
\begin{equation}
 E_c/E_{\rm{vdW}} \approx c_c \left[ L_R(L_R +1)\right]^{3/2} , 
\label{eq:escal2}
\end{equation}
where $c_c = 1.5$ and 
 $E_{\rm{vdW}} = 4 \hbar^3/ (m^{3/2} C_6^{1/2}) $ is the van der Waals energy.
Reading off $E_c (L_R)$  from our numerical results in Fig. \ref{fig:phL} as the first maximum of $|\phi_d (\sqrt{ E_b m/ 3})|^2$ we find 
that the data points are well described by Eq.(\ref{eq:escal2}) when we use  
$c_c = 2.12$, see inset of Fig. \ref{fig:phL}. This validates our simple interpretation of the 
angular momentum suppression. 

\section{Discussion}
\label{sec:discussion}
\begin{figure}[t]
	\centering
	\resizebox{0.45\textwidth}{!}{\includegraphics{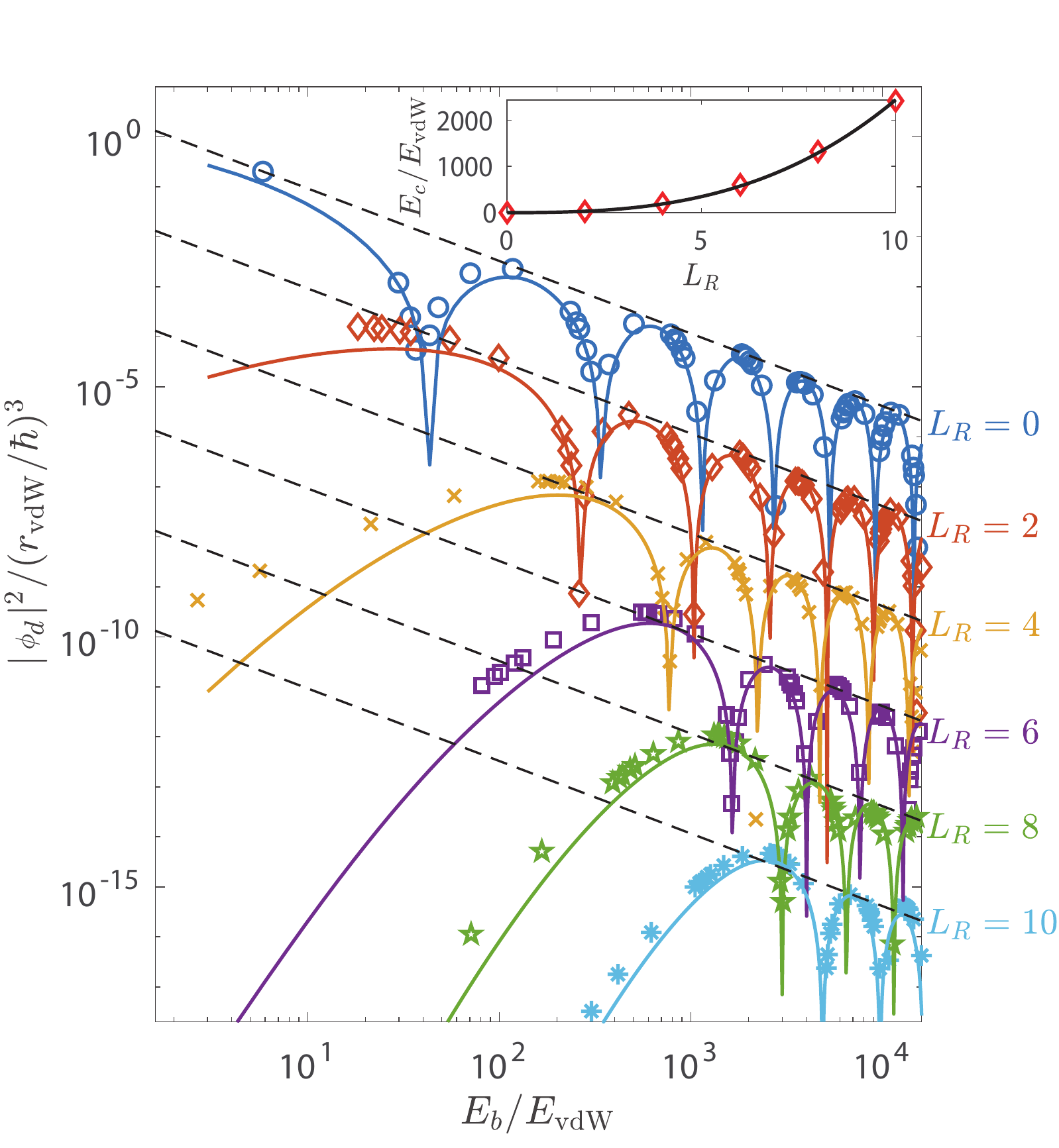} }
	\caption{\label{fig:phL} 
  	Powerlaw scaling of $|\phi_d(\sqrt{E_b\,m/3})|^2 \propto L_3(E_b)/\sqrt{E_b}$  for various rotational angular momenta $L_R$. The data points are a full calculation for the Lennard-Jones potential  $(-C_6/r^6)(1-\lambda_6^6/r^6)$.
 As in Fig.\ref{fig:phi_decay}  and Section \ref{sec:perturbative},
 we  have slightly varied $\lambda_6$ over seven different values while keeping the number of bound states in the potential constant.
  This helps to map out $|\phi_d(\sqrt{E_b\,m/3}) |^2$ in detail.   Therefore, although there are only 8 bound states in the shown energy range we obtain 55 data points.
  The solid lines are calculations
 for the Lennard-Jones potential using Eq. (\ref{eq:aphi}). 
 The dashed lines show the $E_b^{-1.45}$ energy scaling for $|\phi_d|^2$, as determind by a fit to the upper envelope of the data set.
  For better visibility the data for $L_R>0$ are shifted in vertical direction by multiplying them with $10^{-L_R}$.    
  We define $E_c$ to be the energy at which  the curve $|\phi_d|^2$ (for a given $L_R$) has its first maximum (coming from low energy).
   The inset shows $E_c$  as a function of $L_R$ (diamonds).  The data are well described by Eq. (\ref{eq:escal2}) (solid line).
  }
\end{figure}

We now compare and discuss the results of our theoretical and experimental approaches. 
The suppression effect for large $L_R$ and small $E_b$, which is so clearly visible in Fig.$\:$\ref{fig:phL}
is not so obvious  in Fig.$\:$\ref{fig3} where we present our experimental data and our full coupled channel calculations. 
A small suppression effect might only  be recognizable for the state $\vib = -2, L_R=4$ in Fig.$\:$\ref{fig3}. 
 In practice, the observation of suppressed low-energy high-$L_R$ molecular signals can be hampered by various issues.
 By accident it can occur that  no weakly-bound molecular level  with $E_b < E_c$
 exists for a given rotational angular momentum $L_R$. 
 In fact, quantum defect theory predicts 
 for a van der Waals potential
 that  if the most weakly-bound state for a partial wave  $L_R$ is not close to threshold, then 
 also the most weakly-bound state for the partial wave
 $L_R + 4$ will not \cite{Chin2010}.
 Alternatively, the level can be overlooked experimentally, if its signal is too weak.
 It will be overlooked theoretically if the level has a vibrational quantum number beyond the limits of the model potential. 
It should be, however, clear that the suppression mechanism must exist. Indeed, in a recent experiment on
 diatomic molecular reactions, a similar suppression mechanism has been identified \cite{Liu2021}.

When comparing  Fig.$\:$\ref{fig:phL} with  Fig.$\:$\ref{fig3} it is evident 
that the distinct drops of  $|\phi_d (\sqrt{ E_b m/ 3})|^2$ in Fig.$\:$\ref{fig:phL} do not clearly
appear in Fig.$\:$\ref{fig3}. There are several possible explanations for this. First, the data sampling
in  Fig.$\:$\ref{fig3}   is seven times smaller than for Fig.$\:$\ref{fig:phL}. Therefore, it is likely that 
 a narrow drop is not encountered in Fig.$\:$\ref{fig3}.  Second, the calculations in  Fig.$\:$\ref{fig:phL} correspond to the leading order of an expansion. Including higher orders might wash out the sudden drops, as other pathways for the molecular formation can be taken.  

 Concerning the full  model and the experiment, we expect that the scaling
 exponent $\alpha$ of the  $ E_b^{-\alpha}$ scaling law is prone to  changes for deeper binding energies than the ones considered here.
 As recently discussed in \cite{Haze2022}, for $^{87}$Rb and $E_b>150\:\textrm{GHz}\times h$ the spin conservation  propensity rule which allows for working with a single spin channel should break down, affecting the scaling law. 
In addition, the short-range three-body interaction, which is ignored so far in our treatment, 
 should play an increasingly important role when forming more tightly bound molecular states. 
 Recent work on three-body recombination of hydrogen \cite{Yuen2020} has already found evidence for this,
 as the Jahn-Teller effect  substantially enhances recombination rates into tightly bound molecular states.

\section{Summary and Outlook}
\label{sec:outlook}

To summarize, we have experimentally and theoretically investigated how the three-body recombination 
of an ultracold gas scales with the molecular binding energy $E_b$, 
detecting bound levels from 0.02 to 77  GHz $\times h$, thus 
spanning an energy range of more than three orders of magnitude.
This became possible by applying improved experimental schemes for the state-to-state detection of molecules and 
by carrying out large scale numerical calculations.
Besides these numerical calculations an analytical perturbative model was developed which gives deep physical insights into the recombination process and can explain the observed scaling law. In particular, the perturbative model shows that to a large part the scaling law can be extracted from two-body quantities such as the molecular wave function.
Our experimental and theoretical approaches show that the three-body recombination 
exhibits a propensity towards weakly bound product molecules. The recombination rate 
 follows a $E_b^{-\alpha}$ scaling law 
 where  $\alpha$ is in the vicinity of 1.
Remarkably, we find that  this scaling law is quite universal as it should hold for a range of different potentials such as the
Morse potential, potentials of type $-C_n/r^n$ with $n = 3, 4, 6$, as well as the contact potential. 
In addition, apart from a centrifugal barrier suppression at low enough binding energies, our results indicate that
the three-body recombination populates molecular quantum states with different rotational angular momenta quite evenly, within about a factor of two. 

In the future it will be interesting to explore how the scaling law evolves for
 deeper binding energies and what physical mechanisms will lead to its breakdown. 
On the experimental side, the detection sensitivity must be enhanced and the spectroscopic data will be expanded for reliable quantum state identification. On the theory side,  short-range three-body interactions which are ignored so far in our treatment will be taken into account.  
 
Moreover, it will be insightful to explore how deviations  of individual reaction channels from the $E_b^{-\alpha}$ scaling law can be explained  on a microscopic level, e.g. as  interference effects of collision pathways.
In fact, our perturbative calculations already produce 
tell-tale  oscillations
and it will be interesting whether we can match up these oscillations with the ones from the hyperspherical approach. This might give deeper insights into the reaction process. 

We expect that our results on the scaling of the reaction rate with energy  are not restricted to the recombination process of neutral atoms alone, but they can also be applied to other systems and processes. For example, these systems could involve molecules or ions as collision partners and they might also comprise a range of collisional relaxation processes. 
We expect these process rates  to be governed by a  $E_b^{-\alpha}$ scaling law, where $\alpha$ should always be in the vicinity of unity.

\section*{Acknowledgments}
This work was financed by the Baden-W\"{u}rttemberg Stiftung through the Internationale Spitzenforschung program (Contract No. BWST ISF2017-061) and by
the German Research Foundation (DFG, Deutsche Forschungsgemeinschaft) within Contract No. 399903135. We acknowledge support from bwForCluster JUSTUS 2 for
high performance computing. J. P. D. also acknowledges partial support from the U.S. National Science Foundation
(PHY-2012125) and NASA/JPL (1502690).

\appendix%

\section{The (1,1) REMPI }
\label{appendix:REMPI}

Our (1,1) REMPI scheme consists of two excitation steps which are described in the following.
  The overall ionization efficiency $\eta$ is given by the product of the efficiencies of the first and the second (1,1) REMPI step. We achieve $\eta \approx 4.8\times10^{-3}$ (see also Appendix \ref{appendix:Conversion}) for all ($\vib,L_R$) product molecules that we probe in the experiment. This is about an order of magnitude larger as compared to our previous work \cite{Wolf2017, Wolf2019}.

\subsection{First REMPI step}
\label{sec:REMPI1}
 The first REMPI step resonantly excites a  molecule towards the intermediate level $\vib ' =66, \, J'$  of the state $A^1\Sigma_u^+$ using a wavelength of about 1065$\:$nm \cite{Deiss2015,Drozdova2013}.
For this excitation we use a cw external-cavity diode laser with a short-term linewidth of about  $100\:\text{kHz}$.
It has a waist ($1/e^2$-radius) of $\approx 280\:\upmu \text{m}$ and an intensity of $\approx 12\:\text{W}\,\text{cm}^{-2}$ at the position of the atomic cloud.
The laser is frequency-stabilized to a wavelength meter achieving a shot-to-shot and long-term stability of a few megahertz. The laser beam polarization has an angle of
about 45$^\circ$ with respect to the $B$-field and can therefore drive  $\sigma$- and $\pi$-transitions.

\begin{figure}[t]
	\includegraphics[width=\columnwidth]{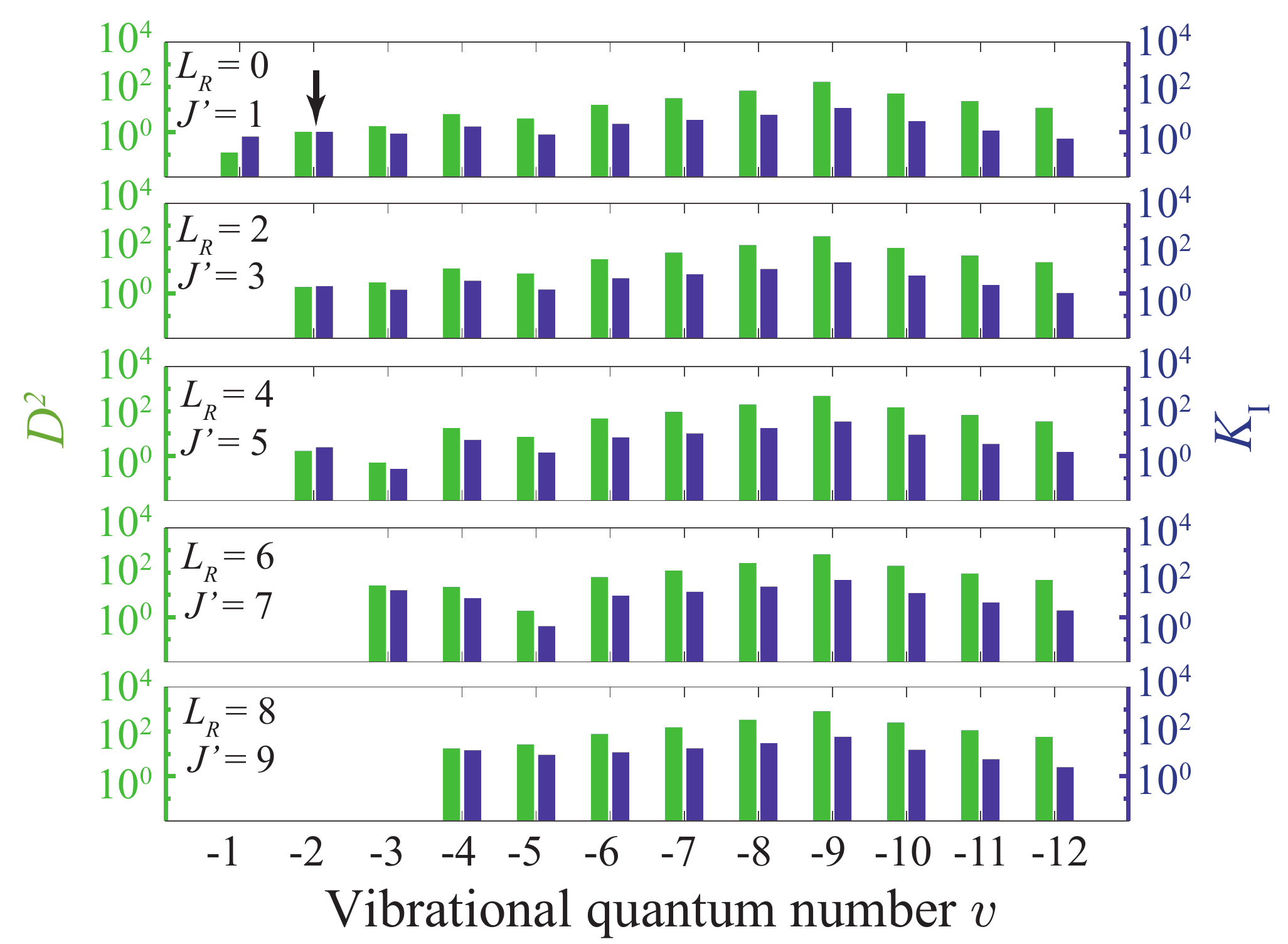}
	\caption{ Calculations for the squared reduced transition electric dipole moment $D^2$  and relative transition probability $K_I\propto D^2f_D / E_b^{1/2}$ for the transitions from the product states ($\vib,L_R$) towards the intermediate states $J', \, \vib ' =66$ of $A^1\Sigma_u^+$. Both, $D^2$ and $K_I$ are normalized by a global factor, respectively, so that their values for the state $\vib = -2, \, L_R = 0$ are 1 (see arrow). The plot is logarithmic.
	}
	\label{fig:App4}
\end{figure}

We estimate that the first REMPI step is saturated for the molecular states considered in this work. This means a molecule is resonantly excited to the intermediate level $J'$  with nearly unit probability when probed.  In order to derive this we consider the following quantities: The transition electric dipole moment, the limited time to optically excite the product molecule, and a detuning due to the Doppler shift. In Fig.$\:$\ref{fig:App4} we show calculated squared reduced transition electric dipole moments $D^2$ for transitions from ($\vib, \, L_R$) product states towards the vibrational level $\vib ' =66$ within $A^1\Sigma_u^+$ (green columns). For convenience, $D^2$ is normalized by a global factor so that its value for ($\vib = -2, L_R = 0$) equals to 1. As a general pattern,  $D^2$ increases with binding energy in the range between 
the threshold and the vibrational quantum number $\vib=-9$. This is, however, partially compensated by the following kinetic effects. In the formation of the molecule by three-body recombination the binding energy is converted into kinetic energy of the products.
Neglecting the energy of the ultracold atoms, the velocity $\vib_\mathrm{Rb2}$ of the molecule is $\propto E_b^{1/2}$. This velocity has two effects. First, it limits the time scale $\propto 1/E_b^{1/2}$ for the molecule to be located in the detection region, as determined by the size of the REMPI laser beams. Second, the velocity will on average lead to Doppler-broadening and to a reduction of the on-resonance photoexcitation rate by a factor given by
$f_D =  (2v_\mathrm{Rb2}/ (\lambda \gamma))^{-1}  \arctan (2v_\mathrm{Rb2}/ (\lambda \gamma)) $, where $\lambda$ is the transition wavelength and $\gamma \approx 2\pi\times15\text{MHz}$ is the linewidth of the excited state. Therefore, the optical excitation probability $K_I$ of the molecules approximately scales as $K_I \propto D^2 f_D /E_b^{1/2}  $.
In Fig.$\:$\ref{fig:App4}  we plot $K_I$ which is also normalized so that for $(\vib = -2, L_R = 0)$ it equals to 1 (blue columns). Again, down to $\vib = -9$ there  is a tendency that the ionization efficiency increases for increasing binding energy. In \cite{Wolf2017} we have found that for the given parameters of the probe laser beam the transitions from $\vib=-2,\, L_R=0, \, 2$ towards the intermediate state are driven in a strongly saturated regime. Since all transitions for the product molecules in Fig.$\:$\ref{fig:App4} have a $K_I$ close or larger than the ones for $\vib=-2, \, L_R=0, \, 2$, we can expect saturation of the first step of the (1,1) REMPI for all considered product levels.

\subsection{Second REMPI step}
\label{sec:REMPI2}
\begin{figure}[t]
	\includegraphics[width=\columnwidth]{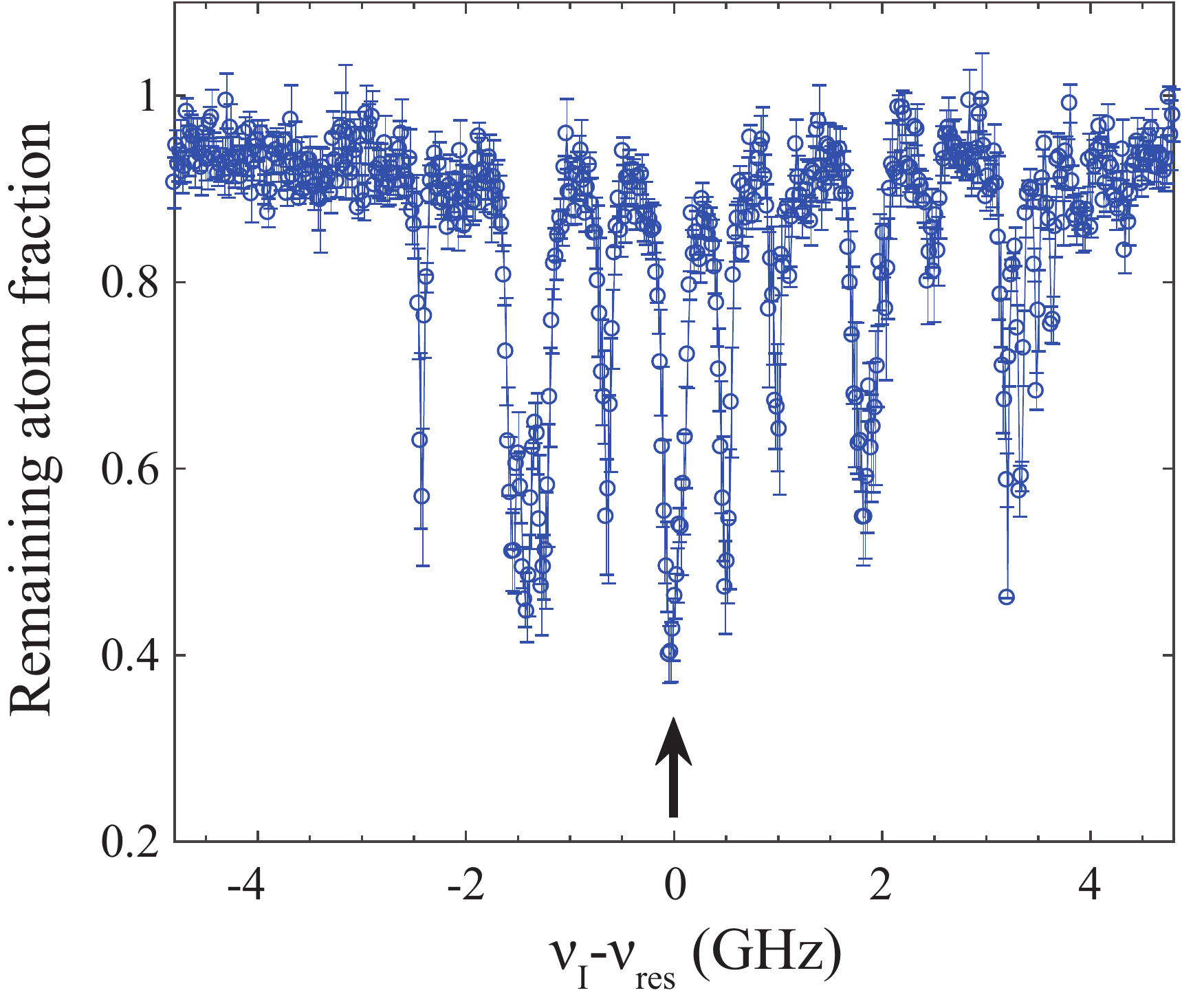}
	\caption{Resonance spectrum when scanning the frequency $\nu_I$ of the ionization laser, starting from the intermediate level $A^1\Sigma_u^+,\, \vib ' =66, \, J'=1$. Here, produced ions are detected by induced loss in an atom cloud (see Appendix \ref{appendix:counting}). The strongest line is marked by an arrow and has the center frequency $\nu_\mathrm{res}=551422.660\:\textrm{GHz}$.
	}
	\label{App_fig4}
\end{figure}

For the second step of the REMPI we use the ionization laser at 544 nm to resonantly drive a transition from the intermediate state to a probably autoionizing molecular Rb$_2$ level [see Fig.$\:$ \ref{fig2}(a)].
The laser is a cw, frequency-doubled OPO system from H\"{u}bner Photonics. It has  a short-term linewidth on the order of $1\:\text{MHz}$. At the position of the atomic cloud the beam waist  is $240\:\upmu \text{m}$ and we typically work with an intensity of $110\:\text{W}\,\text{cm}^{-2}$.
As the probe laser, it is frequency-stabilized to a wavelength meter achieving a shot-to-shot and long-term stability of a few megahertz. The laser polarization is at an angle of about 45$^\circ$ with respect to the $B$-field and can therefore drive  $\sigma$- and $\pi$-transitions.
When  the excited Rb$_2$ molecule autoionizes it produces a deeply bound Rb$_2^+$ molecular ion. Figure \ref{App_fig4} shows resonance lines when scanning the frequency $\nu_I$ of the ionization laser and starting from the intermediate state $A^1\Sigma_u^+, \, \vib ' =66, \, J'=1$ which has been populated via photoassociation. These lines are spectroscopically not yet assigned. For REMPI via $J'=1$ we use the resonance line centered at $\nu_I=\nu_\mathrm{res}=551422.660\:\textrm{GHz}$ (marked by an arrow).

Regarding the intermediate states with $J'>1$, we carry out similar spectroscopy as for Fig.$\:$\ref{App_fig4} and identify the strongest resonance, respectively, which is then used for REMPI. Since photoassociation cannot produce a molecule with $J'>1$ in our cold sample, we instead populate such a state by resonantly exciting suitable product molecules ($\vib,L_R$) after three-body recombination with the probe laser. Table \ref{tab:REMPIpaths} lists the optimal ionization laser frequencies for various rotational levels $J'$ of the intermediate state $\vib '  = 66$. From additional spectroscopic measurements we extract a  rotational constant $B_{\vib '=66 } = 443(2)\:\text{MHz}\times h$,  in agreement with the value reported in \cite{Drozdova2013}. For completeness, we present in Table \ref{tab:REMPIpaths} also the measured level energies $E_{J'}$ for the various detected rotational states  $J'$ within $\vib'=66$ of $A^1\Sigma_u^+$.

\begin{table}
	\begin{tabular}{c  c  c }
		\hline
		\hline
		$J'$ & $E_{J'}$ & $\nu_\mathrm{res}$ \\
		  & $\textrm{GHz}\times h$ & $\textrm{GHz}$ \\
		\hline
		1  \, \, & 281,445.045  & \, \, 551,422.66 \\
		3  \, \, & 281,449.481 & \, \, 551,420.70 \\
		5 \, \, & 281,457.442 & \, \, 551,419.30 \\
		7 \, \, & 281,468.987 & \, \, 551,423.10 \\
		9  \, \, & 281,484.065 & \, \, 551,421.60 \\
		\hline
		\hline
	\end{tabular}
	\caption{REMPI paths. 
 Column two lists the energies $E_{J'}$ of the intermediate levels $J'$ 
in the state $\vib ' =66$ of $A^1\Sigma_u^+$,
with respect to the $(5s, f=1)+(5s,f=1)$ atomic asymptote.  
 In the third column we list the frequencies $\nu_\mathrm{res}$ of the ionization laser which are used in the (1,1) REMPI scheme for the second transition starting from the intermediate levels $J' = 1, 3, 5, 7, 9$  in the state $\vib ' =66$ of $A^1\Sigma_u^+$. 
	\label{tab:REMPIpaths}
	}
\end{table}

\begin{figure}[t]
	\includegraphics[width=\columnwidth]{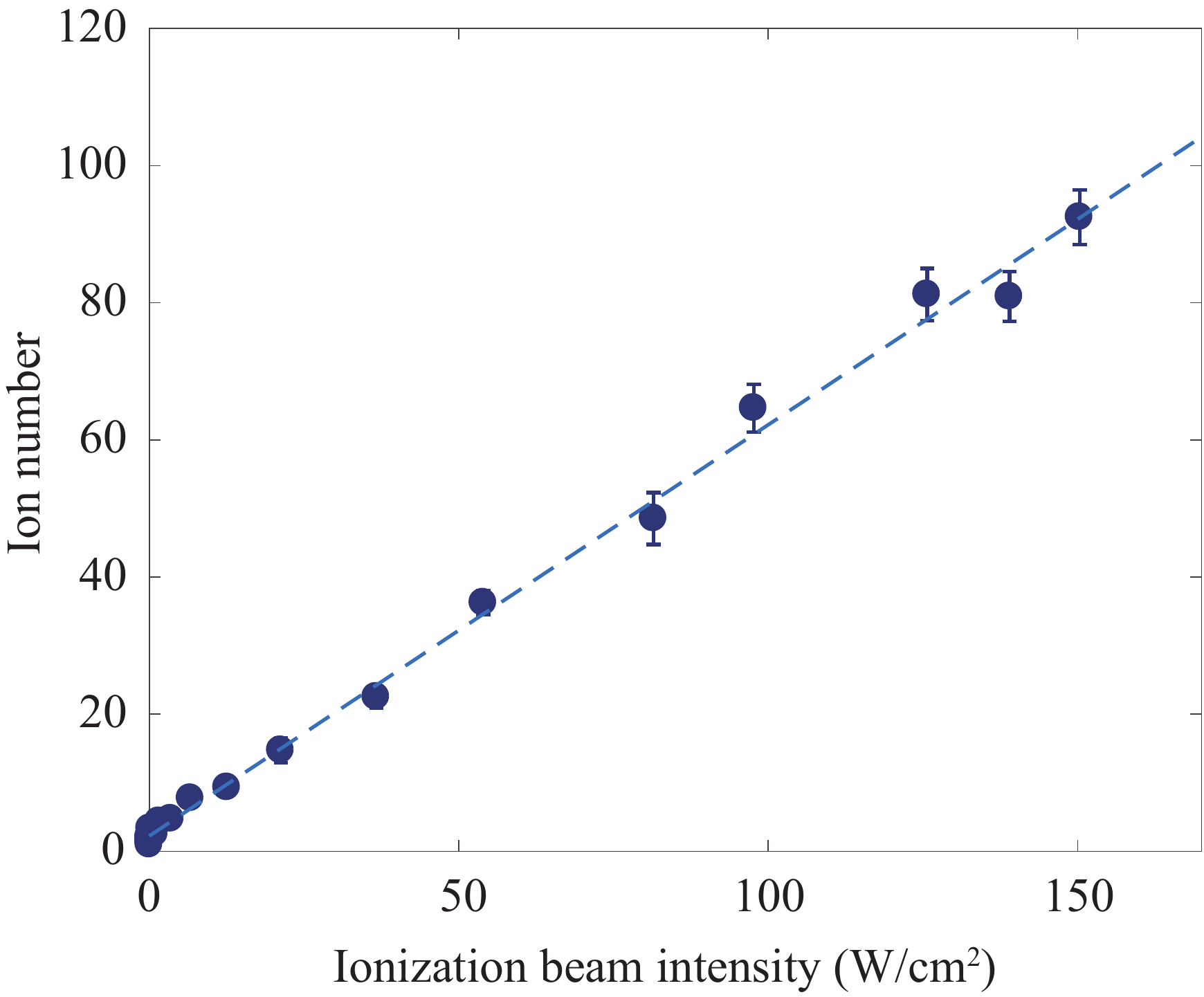}
	\caption{Ion number as a function of the intensity of the ionization laser beam in the (1,1) REMPI scheme. These data are obtained for ($\vib=-2,L_R=2$) molecules which are ionized via the intermediate state $J'=1$. Here, $\nu_I=551,422.66\:\textrm{GHz}$, see Table \ref{tab:REMPIpaths}. The dashed blue line is a linear fit to the data.
	}
	\label{fig:App2}
\end{figure}

The second REMPI step is generally not saturated in our experiment. This is shown in Fig.$\:$\ref{fig:App2} for the case of initial ($\vib=-2,L_R=2$) molecules and the intermediate level $J' = 1$, $\vib ' =66$ of $A^1\Sigma_u^+$. The detected number of ions increases linearly with laser intensity. We have, however, verified that ionization of a given initial state via different intermediate rotational states $J' = L_R \pm 1$  provides similar ion signal strengths. Furthermore, the reduction $f_D$ due to the Doppler effect (as  discussed for the REMPI step 1) should be negligible here, since 
the linewidth $\gamma$ is 
about
$ 200\:\textrm{MHz}$ according to Fig.$\:$\ref{App_fig4}. We note that  $\gamma$ corresponds to an approximate measure of the autoionization width.

\section{Counting REMPI ions via atom loss}
\label{appendix:counting}

\begin{figure}[t]
	\includegraphics[width=\columnwidth]{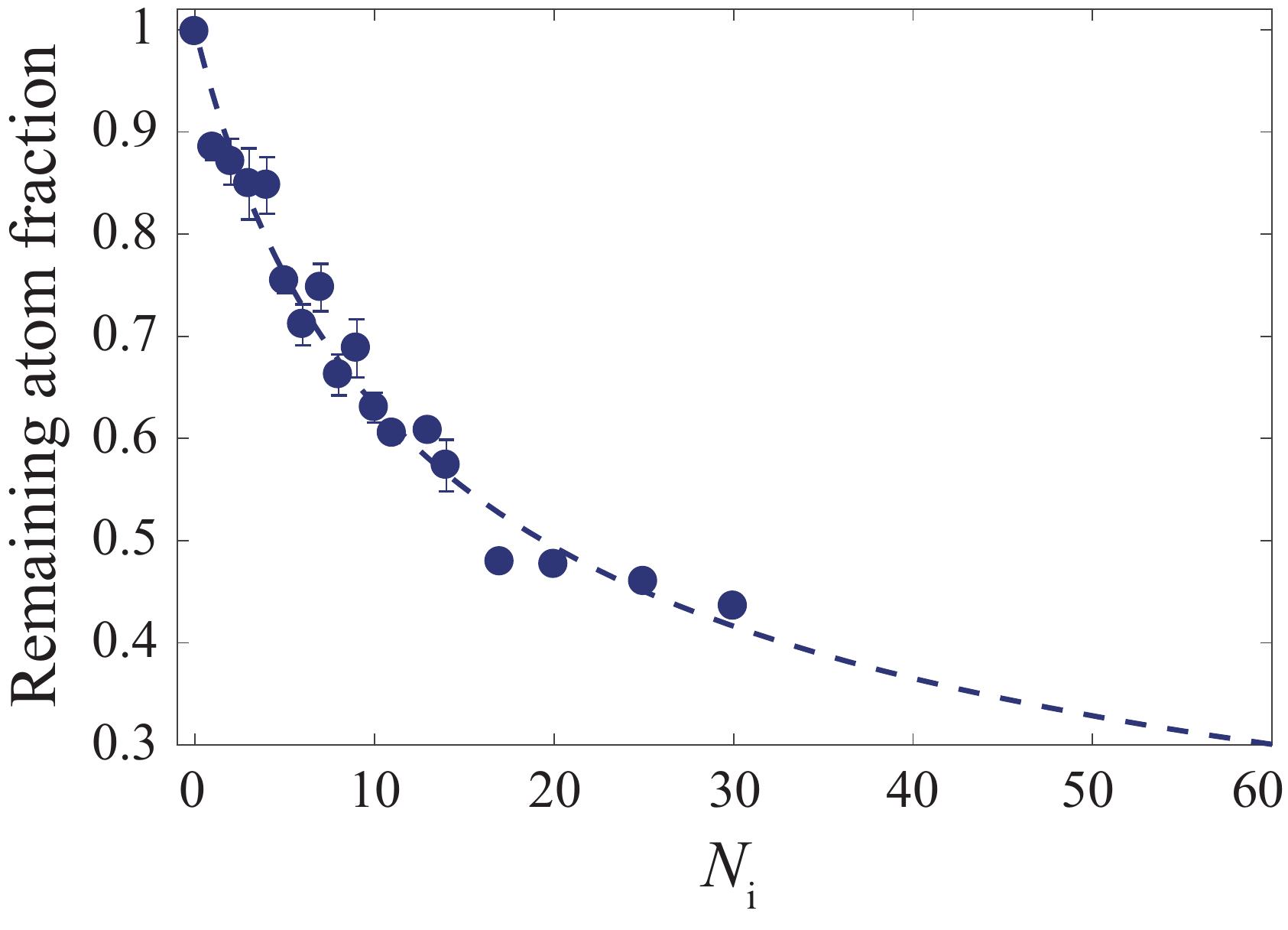}
	\caption{Remaining atom fraction as a function of the initially prepared  number  of ions $N_i$ (Rb$^+$ and Rb$_2^+$)  after an interaction time of $500\:\textrm{ms}$. The dashed line is a fit of the function $ (\frac{N_\text{i}}{c_1}+1)^{c_2}$ to the data (see text).
	}
	\label{fig:App3}
\end{figure}

We detect and count ions in the Paul trap via atom loss which the ions inflict on neutral atoms due to elastic atom-ion collisions. The basic method has been developed in previous work \cite{Haerter2013, Haerter2013b, Wolf2017}. Here, we use a modified scheme, which is described in the following.

{\it Ion-inflicted atom loss:} After the 500$\:$ms phase of three-body recombination, the REMPI lasers are switched off and the optical lattice trap with the atom cloud is adiabatically moved over a distance of $50\:\mu\textrm{m}$  to the center of the Paul trap, in order to immerse the ions
which have been produced during the 
500$\:$ms time
into the atom cloud. At the same time the optical lattice trap is adiabatically converted into a crossed dipole trap, by turning off one of the 1D lattice beams.
The trap frequencies of the crossed optical dipole trap are $\omega_{x,\,y,\,z}=2\pi \times (23,\,180,\,178)\:\text{Hz}$, where $z$ represents the vertical direction. In this trap the atom cloud is Gaussian-shaped with widths of $\sigma_{x,\,y,\,z} \approx (58.6,\,7.5,\,7.5)\:\upmu\text{m}$ and it still consists of about $4\times10^6$ atoms, corresponding to a peak atomic density of about $n_0=0.9\times 10^{14}\:\text{cm}^{-3}$. In the Paul trap the ions have typical kinetic energies on the order of $\text{mK}\times k_B$ or larger as a result of, e.g., excess micromotion. Therefore, in elastic atom-ion collisions one atom after the next is kicked out of the comparatively shallow ($9\:\mu\textrm{K}\times k_B$) optical dipole trap while the ions remain confined in the 2.5$\:$eV deep Paul trap.

{\it Converting atom loss signals into ion numbers:} In order to extract ion numbers from atom loss signals we carry out the following, independent calibration. We start using a configuration where the ion trap and the optical dipole trap are spatially separated from each other so that atoms and ions cannot collide. We prepare a well-defined number of laser-cooled $^{138}$Ba$^+$ ions forming an ion crystal in the Paul trap. Using fluorescence imaging the number of ions in the crystal is counted. In parallel, we prepare a dense Rb atom cloud in the optical dipole trap. Afterwards, the centers of the ion trap and the optical dipole trap are quickly overlaid so that the Ba$^+$ ions are immersed into the atom cloud, where they undergo chemical reactions. This leads almost exclusively to the formation of Rb$^+$ and   Rb$_2^+$ ions  \cite{Mohammadi2021}. The number of ions remains constant because of the large depth of the Paul trap. Subsequently the traps are again separated from each other and a new atom cloud is prepared in the optical dipole trap, while the ions remain confined in the Paul trap. The properties of this new atom cloud are adjusted to match the cloud that was used for atom loss measurements as described in the previous paragraph. Once again the ions are immersed into the atom cloud by overlapping the traps. After an atom-ion interaction time of $500\:\text{ms}$ the remaining number of atoms is measured. In Fig.$\:$\ref{fig:App3} we show data for different numbers $N_i$ of prepared ions. The blue dashed line represents a fit using the empirical function $(\frac{N_\text{i}}{c_1}+1)^{c_2}$, where the fit parameters are $c_1=8.6$ and $c_2=-1.6$. This function serves as reference to convert measured atom loss in the three-body recombination experiment into ion numbers.\\

\section{Conversion of REMPI ion numbers to molecule numbers and $L_3$ rate coefficients}
\label{appendix:Conversion}

As discussed in Appendix \ref{appendix:REMPI}, the number of formed molecules in a final channel is given by the measured ion number $N_\mathrm{Ion}$ after REMPI divided by a global ionization efficiency factor $\eta$. Similarly,  the   $L_3(\vib,L_R)$ rate coefficients are  given by the measured ion numbers multiplied with a proportionality factor $\kappa$. We note that each three-body recombination experiment has a run time of 500 ms, and after this time 
the atomic density of the sample   has dropped by
13\%  due to collisional and reactive loss. Taking this loss into account we 
estimate the ion number $N_\mathrm{Ion}^* = N_\mathrm{Ion} \times 1.22 $ for the case that the density stayed constant. 
 To determine the proportionality factor $\kappa$, we sum over all channels according to $ \kappa  \sum_{\vib, L_R}  N_\mathrm{Ion}^*(\vib, L_R) =  \sum_{\vib, L_R} L_3(\vib, L_R) = L_3$, where $L_3(\vib, L_R)$ and $L_3$ are the calculated partial and total recombination rate constants, respectively. We obtained 
  $\kappa=1.92\times10^{-32}\:\textrm{cm}^6\textrm{s}^{-1}$.
From an additional measurement of the initial atom number and the temperature of the atom cloud and using the
molecular production rate
\begin{equation}
\dot{M} = \frac{L_3}{3\sqrt{27}}\left(\frac{m \bar{\omega}^2}{2\pi k_B}\right)^3 \frac{N^3}{T^3}. 
\label{eq:TBR}
\end{equation}
we determined $\eta $ to be $\eta \approx 
4.8\times10^{-3}$.
Here, $M$ is the number of produced molecules, $m$ is the atomic mass, and 
$\bar{\omega} = (\omega_x \omega_y \omega_z)^{1/3}$ is the geometric mean of the
trapping frequencies. 

\section{Boosting the molecule signals}
\label{appendix:boost}

Compared to our previous work \cite{Wolf2017, Wolf2019} the product state signals and the sensitivity were boosted by a factor of $\approx 25$ as a result of two improvement steps.

\begin{figure}[t]
	\includegraphics[width=\columnwidth]{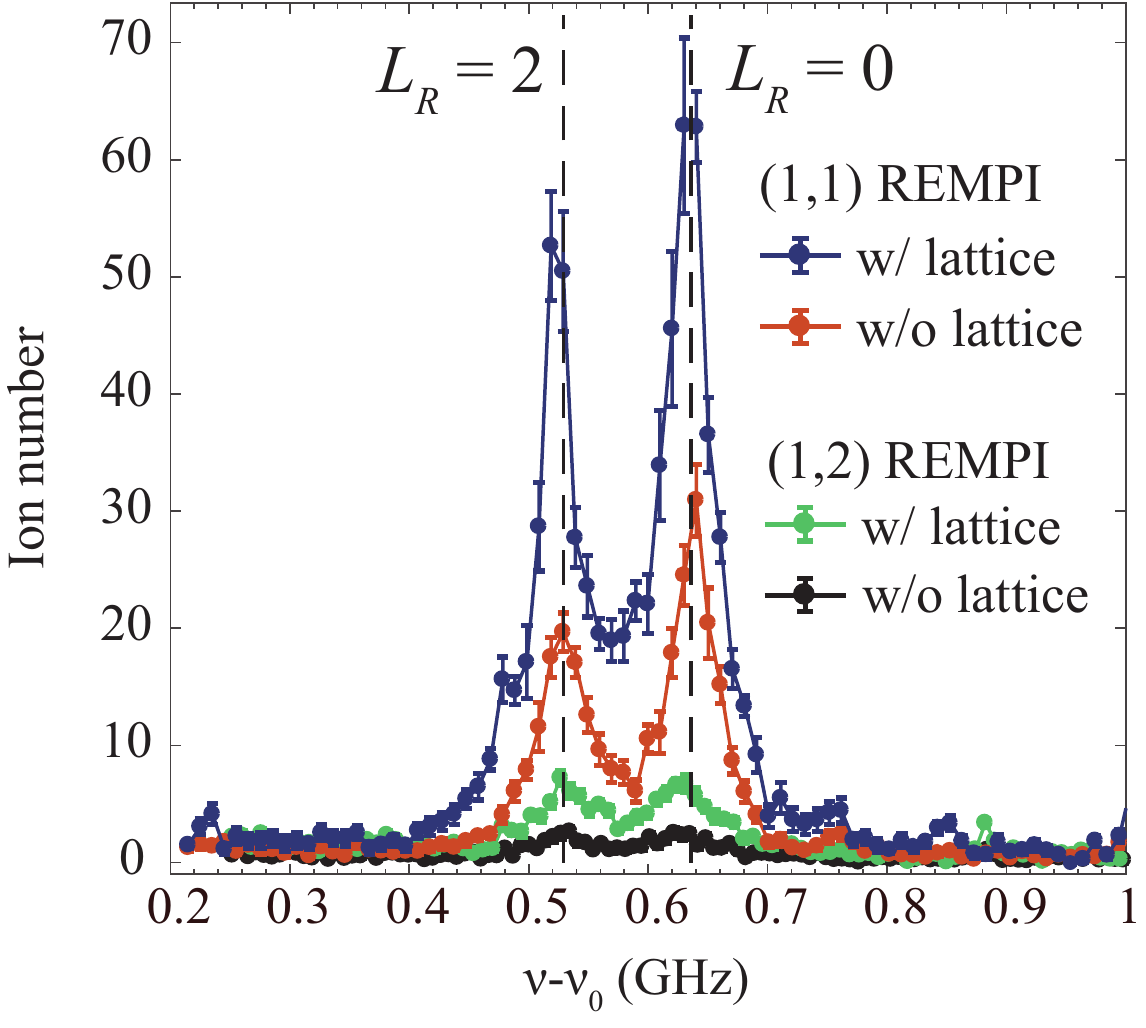}
	\caption{Boost of the molecular detection signal. Shown are measured ion numbers after REMPI of ($\vib=-2, L_R=0,2$) product molecules as a function of the probe laser frequency $\nu$.
		$\nu_0=281445.045\:\textrm{GHz}$ is the resonance frequency for photoassociation towards the $J' = 1$ intermediate state.	
		The different colors indicate different measurement configurations as detailed in the legend.
	}
	\label{fig:App6}
\end{figure}

The first improvement step is working with an optical lattice trap instead of a  plain crossed dipole trap. This increases the atomic density $n$ and therefore improves the signal compared to the background since the three-body recombination rate scales non-linearly with $n$ as $\dot{N}_{\text{at}}=\int \dot{n}\,\text{d}^3r=-L_3\int n^3\,\text{d}^3r$, whereas background signals scale less strongly with
$n$. In total, switching  to the optical lattice configuration in our set up increased  the signal by a factor of 2.5.
This improvement is shown in Fig.$\:$\ref{fig:App6} for the detection signals of the $\vib=-2, \, L_R=0, \, 2$ product molecules when using the intermediate state $J' = 1$. We plot the ion number (i.e. the molecular (1,2) REMPI signal) as a function of the probe laser frequency $\nu$ for different experimental settings. The black data points are the signals for the settings in \cite{Wolf2017}. For the green data points we have replaced the
crossed dipole trap as used in \cite{Wolf2017} by the optical lattice.

The second improvement step is an enhancement of the REMPI efficiency. In \cite{Wolf2017, Wolf2019}
 only the first REMPI step in a (1,2) REMPI configuration was resonantly driven. In our new (1,1) scheme [see Fig.$\:$\ref{fig2}(a) and Appendix \ref{appendix:REMPI}] both REMPI steps are resonantly driven and the last excitation step is resulting in a
 molecular state which is probably autoionizing. Switching to the (1,1) REMPI increased the signals by a factor of 10. For comparison, in Fig.$\:$\ref{fig:App6}, the blue and orange data give the signals for (1,1) REMPI with and without lattice, respectively.

In order to  optimize REMPI we have also experimentally tested several vibrational levels around $\vib ' =66$ within $A^1\Sigma_u^+$ as intermediate level, but obtained the best results for $\vib ' =66$ in terms of efficiency and suppressing background signals. The level $ \vib ' =66$  has essentially a simple rotational ladder structure, as
it has unresolved hyperfine splittings of less than $3\:\text{MHz}\times h$ \cite{Deiss2015, Drozdova2013}.

\section{Consistency checks for line assignments}
\label{appendix:ConsistencyCheck}

\begin{figure}[t]
	\includegraphics[width=\columnwidth]{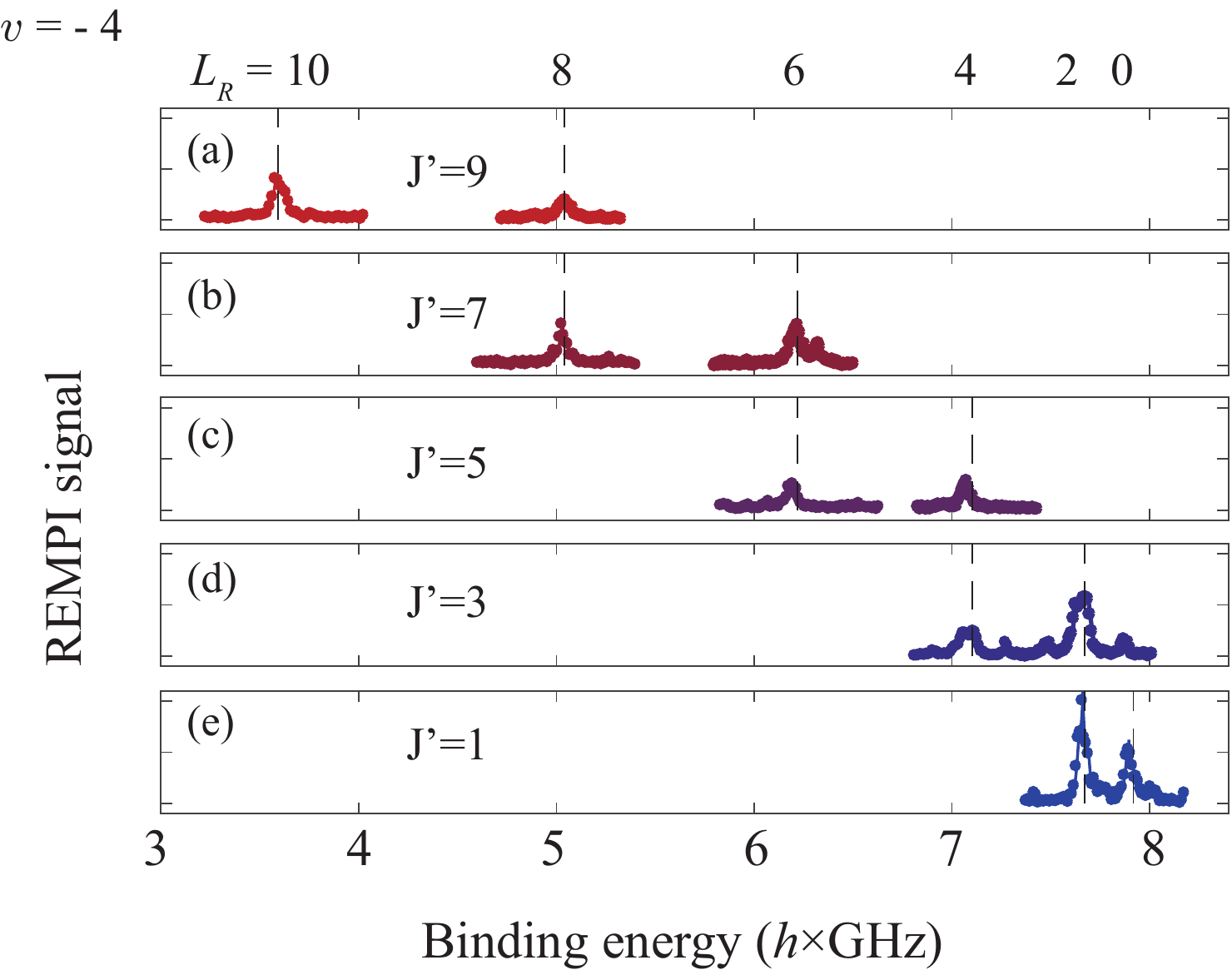}
	\caption{Consistency of assignments of product molecular states to REMPI detection signals. The vertical dashed lines mark the binding energies of the molecules with quantum numbers $\vib=-4,L_R$ as obtained from coupled channel calculations. The rotational quantum number $L_R$ is shown on top of the figure. For each level with $L_R > 0$ we observe two resonance lines, corresponding to REMPI via the intermediate levels $J' = L_R \pm 1$. The resonance lines are plotted as a function of the probe laser frequency and are arranged such that the frequency position of each resonance peak is given by $\nu - \nu_0 - B_{\vib'=66}  \times [J' (J' +  1)-2]$, ideally corresponding to the molecular binding energy.	The rotational constant is $B_{\vib'=66} = 443\:\textrm{MHz}$. $\nu_0$ is the photoassociation resonance frequency towards $J'=1$. In vertical direction the lines are arranged according to $J'$. The match in frequency position of the two measured resonance lines and the calculated binding energy for a level indicates consistency of the assignment.
	}
	\label{fig:App5}
\end{figure}

In general, we observe each molecular level with $L_R > 0$ in terms of two transition lines,
$L_R \rightarrow J' = L_R \pm 1 $. This greatly helps to verify the consistency of the line assignment. As an example, we provide in Fig.$\:$\ref{fig:App5} the collection of REMPI detection signals for $\vib=-4$ molecules with rotational states $L_R$ ranging from $L_R=0$ to $10$.

\section{Three-body model for $^{87}$Rb atoms}
\label{appendix:TBmodelRb87}

Our three-body calculations for $^{87}$Rb atoms were performed using the adiabatic hyperspherical representation \cite{dincao2018JPB,wang2011PRA} where the hyperradius $R$ determines the overall size of the system, while all other degrees of freedom are represented by a set of hyperangles $\Omega$. Within this framework, the three-body adiabatic potentials $U_{\alpha}$ and channel functions $\Phi_{\alpha}$ are determined from the solutions of the hyperangular adiabatic equation:
\begin{align}
\left[\frac{\Lambda^2(\Omega)+15/4}{2\mu R^2}\hbar^2+\sum_{i<j}V(r_{ij})\right]\Phi_{\alpha}(R;\Omega)
=U_{\alpha}(R)\Phi_{\alpha}(R;\Omega),
\end{align}
which contains the hyperangular part of the kinetic energy, via the grand-angular momentum operator $\Lambda^2$, and the three-body reduced mass $\mu=m/\sqrt{3}$. To calculate the three-body recombination rate we solve the hyperradial Schr\"odinger equation,
\begin{align}
&\left[-\frac{\hbar^2}{2\mu}\frac{d^2}{dR^2}+U_{\alpha}(R)\right]F_{\alpha}(R)\nonumber\\
&~~~~~~+\sum_{\alpha'}W_{\alpha\alpha'}(R)F_{\alpha'}(R)=EF_{\alpha}(R),\label{Schro}
\end{align}
where $\alpha$ is an index that labels all necessary quantum numbers to characterize each channel, $E$ is the total energy, and nonadiabatic couplings are given by 
\begin{align}
W_{\alpha\alpha'}(R)=-\frac{\hbar^2}{2\mu}\left(\langle\Phi_{\alpha}|\frac{d^2}{dR^2}\Phi_{\alpha'}\rangle+2\langle\Phi_{\alpha}|\frac{d}{dR}\Phi_{\alpha'}\rangle\frac{d}{dR}\right).
\end{align}
Solving Eq.$\:$(\ref{Schro}) numerically \cite{wang2011PRA} we determine the $S$-matrix and the
recombination rate $L_3$ via
\begin{align}
    L_3=\sum_{f,i}\frac{96\pi^2(2J+1)\hbar}{\mu k_i^4}|S_{fi}|^2=\sum_{f}L_3(f),
\end{align}
where $i$ and $f\equiv(\vib,L_R)$ represent the initial and final states and $L_{3}(f)$ is the corresponding partial recombination rate for a given final state. 

In this present study, the interaction between $^{87}$Rb atoms is modeled by the same  potential used in Ref.$\:$\cite{Wolf2017}, and is given by the Lennard-Jones potential,
\begin{align}
V(r)=-\frac{C_6}{r^6}\left(1-\frac{\lambda^6}{r^6}\right),\label{vpot}
\end{align}
where $C_6=4710.431E_h$a$_0^6$ is the van der Waals dispersion coefficient from Ref.~\cite{Strauss2010}. We adjust the value of $\lambda$ to have different numbers of diatomic bound states supported by the interaction, while still reproducing the value of the scattering length for $^{87}$Rb atoms, $a=100.36a_0$ \cite{Strauss2010}. Our calculations were performed using $\lambda\approx15.5943a_0$, producing 15 $s$-wave ($L_R=0$) molecular bound states, and a total of 240 bound states including higher partial-wave states, $L_R > 0$. 

Our numerical calculations for three-body recombination through the solutions of Eq.~(\ref{Schro}) have included up to 300 hyperspherical channels leading to a total rate converged within a few percent. The calculated total recombination rate constant at 0.8$\:\mu$K (including thermal averaging) is $L_3\approx0.91\times 10^{-29}$ cm$^6/$s.
 We note that there is an unresolved discrepancy between this calculated total three-body recombination rate constant and the corresponding experimental value for
 Rb in the spin state $f = 1, m_f  = -1$. 
 The experimental value is
$L_3 =(4.3 \pm 1.8)\times 10^{-29}\:\text{cm}^6\text{s}^{-1}$,  \cite{Burt1997}.
  This discrepancy, however, does not affect the analysis of the overall scaling behavior of $L_3(\vib,L_R)$ with binding energy.

\section{Interpreting the $1/E_b$ scaling law via a perturbative approach}
\label{appendix:AGS}

The Alt-Grassberger-Sandhas (AGS) equation  is an efficient approach for solving three-particle collision problems \cite{Alt:1967}, and  
for identical particles it reads \cite{Secker:2021,Li:2022}
\begin{align} \label{eq:AGSeqrecom}
U_{0} (E) & = \frac{1}{3}G_0^{-1}(E) \left[1 + P_+ + P_-\right] \nonumber \\
& \phantom{=} + \left[ P_+ + P_- \right] \mathcal{T} (E) G_0(E) U_{0} (E) \, .
\end{align}
 Here, the three-body transition operator $U_{0}$ describes the transition process from the initial  state of three free, noninteracting atoms to product states of a molecule of the atom pair $(a,b)$ plus a free atom $c$. $E$ denotes the total energy of the three-body system,
    $G_0(E)=(E + i\epsilon -H_0)^{-1}$ is the free Green's operator corresponding to the non-interacting three-body Hamiltonian $H_0$, and  $\epsilon$ is a small quantity to shift the energy away from the real axis.  $\mathcal{T}(E)=\int d \mathbf{q}|\mathbf{q}\rangle t(E_{2b})\langle \mathbf{q}|$ represents the generalized two-body transition operator for the interacting pair $(a,b)$  \cite{Secker:2021,Li:2022}. Here,  $t(E_{2b})$ 
  is the pure two-body transition operator at two-body energy
 $E_{2b}=E-\frac{3q^2}{4m}$, $\mathbf{q}$ is the 3-dimensional relative momentum between atom $c$
 and the center of mass of the pair $(a,b)$,
 $m$ is the mass of an atom and $q$ is the absolute value  of $\mathbf{q}$. $P_+=P_{bc}P_{ab}$ and $P_-=P_{ac}P_{ab}$ denote the cyclic and anticyclic permutation operators for the atoms $(a, b, c)$, respectively. The partial three-body recombination rate $L_3(\vib, L_R)$ towards each specific molecular product $d$ is given by \cite{Li:2022}
\begin{equation} \label{eq:K3}
L_3(\vib, L_R)= \frac{12 \pi m}{ \hbar} (2 \pi \hbar)^6 q_{d} |\langle\psi_{f} | U_{0} (E) | \psi_{\rm{in}} \rangle |^2,
\end{equation}
where $| \psi_{\rm{in}}\rangle$ and $|\psi_{f}\rangle$ represent the initial and product states, respectively.  Here, $q_{d}$ denotes the absolute value of the asymptotic momentum of the molecule which is fixed by the molecule binding energy $E_b$ and the total energy $E$ via $\frac{3q_{d}^2}{4m}-E_b=E$. Iteratively plugging Eq.$\:$(\ref{eq:AGSeqrecom}) into its right side, one gets a series expansion $U_{0} (E)=\sum_{n=0}^{\infty}U_{0}^{(n)} (E)$ with $U_{0}^{(n)} (E)= \left\{\left[ P_+ + P_- \right] \mathcal{T} (E) G_0(E)\right\}^{n}\frac{1}{3}G_0^{-1}(E) \left[1 + P_+ + P_-\right]$. We assume that the three-body recombination process can be reasonably well described by the leading order contribution for $U_{0}$ \cite{Li:2022}. Since $U_{0}^{(0)} (E)| \psi_{\rm{in}}\rangle = \frac{1}{3}\left[1 + P_+ + P_-\right](E-H_0)| \psi_{\rm{in}}\rangle=0$ due to energy conservation, $U_{0}^{(0)} (E)$ has no contribution to the three-body recombination rate according to Eq.$\:$(\ref{eq:K3}). Therefore, we approximate $U_{0}(E)$ by $U_{0}^{(1)}(E)$
\begin{equation} \label{U1}
U_{0}(E)   \approx   \left[ P_+ + P_- \right] \mathcal{T}(E)\frac{1}{3}\left[1 + P_+ + P_-\right].
\end{equation}
Since the three-body recombination rate is usually quite energy independent
 in the ultracold regime \cite{Suno:2002}, we take the zero energy limit $E\rightarrow 0$ to simplify the derivation. In this limit, the initial free atom state is $|\psi_{\rm{in}}\rangle=|\mathbf{p}=0,\mathbf{q}= 0\rangle$, where $\mathbf{p}$ describes the relative momentum between atoms $a$ and $b$. We note that, for identical particles, neither the result of $L_3(\vib, L_R)$ nor the derivation procedure associated to this quantity should depend on the choice of pair $(a,b)$. 
 Plugging the expression of Eq.$\:$(\ref{U1}) into Eq.$\:$(\ref{eq:K3}) we obtain
 \begin{eqnarray} \label{K32}
 &L_3&(\vib, L_R) \\ &\approx & \frac{12 \pi m}{ \hbar} (2 \pi \hbar)^6 q_{ d} |\langle\psi_{f} | \left[ P_+ + P_- \right] \mathcal{T}(E)\frac{1}{3}\left[1 + P_+ + P_-\right] | \psi_{\rm{in}} \rangle |^2 \notag \\
 &=&\frac{12 \pi m}{ \hbar} (2 \pi \hbar)^6 q_{ d}  |\langle\psi_{f} | 2P_+ \mathcal{T}(E) |\psi_{\rm{in}} \rangle |^2,
 \end{eqnarray}
 where we have replaced $P_+ + P_-$ by $2P_+$
 because the term of $P_+$ will contribute equally as the term of $P_-$  to $L_3(\vib, L_R)$, \cite{Secker:2021,Glockle:1983}.
 For similar reasons $[1 + P_+ + P_-]/ 3$ is replaced by 1.
  In the plane wave basis, $P_+$ is given by 
  \begin{equation}
  P_+=\int d\mathbf{q}' \int d\mathbf{q}''|\mathbf{p}',\mathbf{q}'\rangle \langle \mathbf{p}'',\mathbf{q}'' |,
  \label{eq:pplus}
  \end{equation}
where
  $\mathbf{p}' = \mathbf{q}''+\frac{1}{2}\mathbf{q}'$ and
  $\mathbf{p}'' = -\mathbf{q}'-\frac{1}{2}\mathbf{q}''$. 
  To derive Eq. (\ref{eq:pplus}), we let the single atom momenta be \{$\mathbf{k}''_a=\mathbf{k}_1$, $\mathbf{k}''_b=\mathbf{k}_2$, $\mathbf{k}''_c=\mathbf{k}_3$\} and 
  by definition
  we have \{$\mathbf{p}''=(\mathbf{k}_2-\mathbf{k}_1)/2$, $\mathbf{q}''=2 \mathbf{k}_3/3-\mathbf{k}_1/3-\mathbf{k}_2/3$\}. The permutation operator $P_+$ changes the atom indices according to  $(a,b,c)\rightarrow(c,a,b)$ and therefore \{$\mathbf{k}''_a=\mathbf{k}_1$, $\mathbf{k}''_b=\mathbf{k}_2$, $\mathbf{k}''_c=\mathbf{k}_3$\} $\rightarrow$ \{$\mathbf{k}'_a=\mathbf{k}_2$, $\mathbf{k}'_b=\mathbf{k}_3$, $\mathbf{k}'_c=\mathbf{k}_1$\}, which leads to \{$\mathbf{p}''=(\mathbf{k}_2-\mathbf{k}_1)/2$, $\mathbf{q}''=2 \mathbf{k}_3/3-\mathbf{k}_1/3-\mathbf{k}_2/3$\} $\rightarrow$ \{$\mathbf{p}'=(\mathbf{k}_3-\mathbf{k}_2)/2$, $\mathbf{q}'=2 \mathbf{k}_1/3-\mathbf{k}_2/3-\mathbf{k}_3/3$\}, or equivalently,
  \{$\mathbf{p}''=-\mathbf{q}''/2-\mathbf{q}'$, $\mathbf{q}''$\} $\rightarrow$ \{$\mathbf{p}'=\mathbf{q}'/2+\mathbf{q}''$, $\mathbf{q}'$\}. It is then straightforward that $P_+=\int d\mathbf{q}' \int d\mathbf{q}''|\mathbf{p}'=\mathbf{q}'/2+\mathbf{q}'',\mathbf{q}'\rangle \langle \mathbf{p}''=-\mathbf{q}''/2-\mathbf{q}',\mathbf{q}'' |$.
  Using the previous expressions for $P_+$ and $\mathcal{T}(E)$, and
  $|\psi_{\rm{in}}\rangle=|\mathbf{p}=\mathbf{0},\mathbf{q=\mathbf{0}}\rangle$,
  we find
  \begin{eqnarray} \label{K33}
  & & \langle\psi_{f} | P_+ \mathcal{T}(E) |\psi_{\rm{in}} \rangle   \\
  & = & \langle\psi_{f} | \int d\mathbf{q}' \int d\mathbf{q}''
  | \mathbf{p}',\mathbf{q}'\rangle \langle \mathbf{p}'',\mathbf{q}''
   |\int d \mathbf{\tilde q}|\mathbf{\tilde q}\rangle t (E-\frac{3 {\tilde q}^2}{4m})\langle \mathbf{\tilde q}|\psi_{\rm{in}} \rangle
   \notag  \\
  & = &  \langle\psi_{f} | \int d\mathbf{q}' \int d\mathbf{q}''
  | \mathbf{p}',\mathbf{q}'\rangle \langle \mathbf{q}'' | 0\rangle    \langle \mathbf{p}'' | t(0) | \mathbf{p} = 0\rangle  \notag  \\
   & = &   \int d\mathbf{q}'
   \langle\psi_{f} | \mathbf{p}',\mathbf{q}'\rangle   \langle \mathbf{p}'' | t(0) | \mathbf{p} = 0\rangle,  \notag
  \end{eqnarray}
  where we have used  $\langle \mathbf{q}' | \mathbf{\tilde q} \rangle = \delta( \mathbf{q}' - \mathbf{\tilde q}  )$, and  in the last line we have $\mathbf{p}'  = \frac{1}{2}\mathbf{q}' $
  and $\mathbf{p}'' = -\mathbf{q}'$.
 We now switch from the plane wave basis to a partial wave basis by using
 \begin{equation} \label{eq:basisswitch}
 |\mathbf{p} \rangle =    |p \rangle \sum_{l,m} Y_{l,m}^* (\mathbf{e_p}) | l, m \rangle,
 \end{equation}
 where  $p = |\mathbf{p}|$, $ \mathbf{e_p} = \mathbf{p} / p$ and the normalization of 
 $|p\rangle$ is given by  $\langle p'|p\rangle=\frac{\delta(p-p')}{p^2}$. 
 The expression $ \langle \mathbf{p}'' | t(0) | \mathbf{p} = 0\rangle$  in the last line of Eq.$\:$(\ref{K33}) can be expressed as
  \begin{eqnarray} \label{K35}
  \langle -\mathbf{q'} | t(0) | \mathbf{p} = 0\rangle
 & = &  \frac{1}{4\pi}\langle q' | t^s (0) | p = 0\rangle,
   \end{eqnarray}
  where $t^{s}$ is the $s$-wave component of the two-body transition operator $t$.
  Here we have used that according to the Wigner threshold law at
   low collision energies only $s$-wave collisions can contribute. Furthermore, we assume that the interaction between two atoms conserves angular momentum.

 Next,  we consider the expression $  \langle\psi_{f} |\mathbf{p}' = \frac{1}{2}\mathbf{q}' ,\mathbf{q}'\rangle  $.
  For $| \psi_{f} \rangle $ we make the Ansatz $ | \psi_{f} \rangle =  | \phi_{d} \rangle | q_d, \hat l, \hat m\rangle $.
  Here,  $ | \phi_{d} \rangle =   | L_R, m_{L_R} \rangle  \,  \int d p \, p^2 \phi_d(p) | p \rangle $ is the internal wave function of the molecule with rotational angular momentum quantum numbers $ L_R, m_{L_R}$.
  $\phi_d(p)$ is normalized via $\int |\phi_d(p)|^2 p^2 dp=1$.
  The state $| q_d, \hat l, \hat  m\rangle  \equiv | q_d \rangle | \hat l, \hat m\rangle $ describes the relative motion between molecule and atom.  It corresponds to a partial wave with rotational quantum numbers $\hat l, \hat m $.
  Next, we  calculate that
    \begin{eqnarray} \label{K36}
   \langle \phi_d |  \frac{1}{2} \mathbf{q'}\rangle  &=& \phi_d^* \left( \frac{1}{2} q' \right)  \, \,   Y^*_{L_R, m_{L_R}} (\mathbf{e_{q'}} )\,, \\
   \langle q_{d},  \hat{l}, \hat{m} |   \mathbf{q'}\rangle  & = &
     \frac{\delta(q_d - q')}{q^2}   \, \,   Y^*_{\hat{l},\hat{m}} (\mathbf{e_{q'}})\,.
\label{K37}
   \end{eqnarray}
   Plugging the results of Eqs.$\:$(\ref{K35}) to (\ref{K37}) into Eq.$\:$(\ref{K33}) and carrying out the integration over $\mathbf{q'}$ we obtain
   \begin{eqnarray} \label{K34}
  & & \langle\psi_{f} | P_+ \mathcal{T}(E) |\psi_{\rm{in}} \rangle   \\
  & = &
  \frac{(-1)^m}{4 \pi}  \phi^*_d (\frac{1}{2}q_d) \,  \langle q_d|t^{s}(0)|p=0\rangle \,   \delta_{L_R,\hat l} \delta_{m_{L_R}, -\hat m} ,
  \notag
  \end{eqnarray}
  where we have used $ \int d\Omega_{q} Y^*_{lm}(\mathbf{e_{q}})   Y^*_{ l'  m'}(\mathbf{e_{q}}) =
   (-1)^m \delta_{l, l'} \delta_{m, - m'}  $. We then define 
   \begin{equation} \label{eq:thd}
   t_{\rm{h}}(q_d)=\langle q_d|t^{s}(0)|p=0\rangle
   \end{equation}
     and use Eq.$\:$(\ref{K34})  to rewrite Eq.$\:$(\ref{K32}) as
\begin{equation} \label{K31}
L_3(\vib, L_R) \approx \frac{12 \pi m}{ \hbar} (2 \pi \hbar)^6 q_d (2 L_R+1) |\frac{1}{2 \pi}\phi_d \left(\frac{1}{2}q_d \right)t_{\rm{h}}(q_d)|^2.
\end{equation}
Here, we have summed over the $2L_R +1$ equal contributions corresponding to the available
$m_d$-channels for a given $L_R$ quantum number.
$t_{\rm{h}}(q_d)$ has momentum $p = 0$ fixed on the energy shell $p^2/m=E_{2b} = 0 $, and is commonly referred to as half-shell $t$-matrix in nuclear physics \cite{Ernst:1973,Hlophe:2013}.

\begin{figure}[t]
	\centering
	\resizebox{0.5\textwidth}{!}{\includegraphics{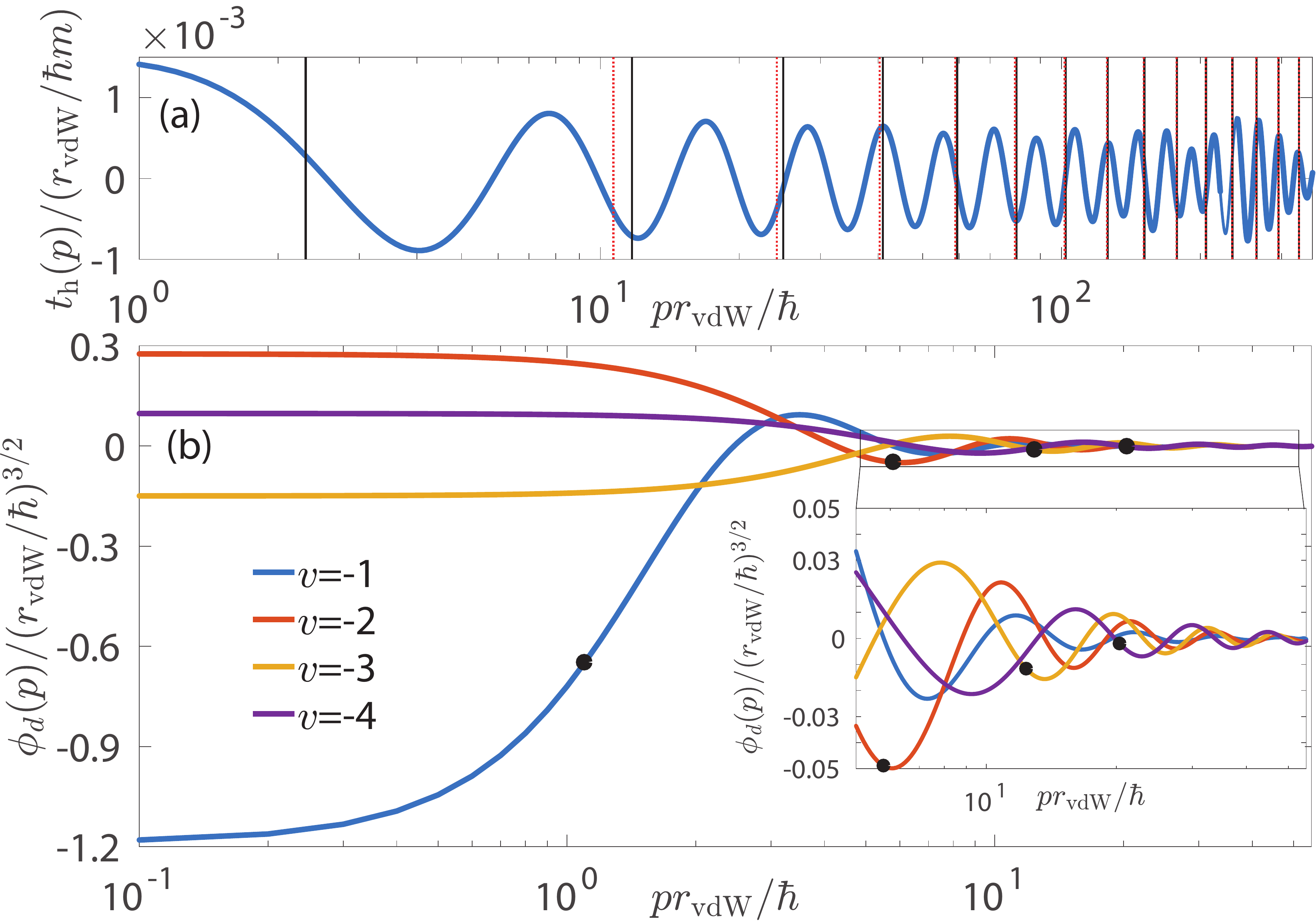}}
	\caption{ (a) Two-body half-shell $t$-matrix $t_{\rm{h}}$ as a function of the relative two-body momentum $p$. The vertical lines indicate the momentum values $2\sqrt{mE_b/3}$ corresponding to $s$-wave (black solid) and $d$-wave (red dotted) molecules where $t_{\rm{h}}$ needs to be evaluated. (b) Momentum space wave functions of the four most shallow $s$-wave molecules. The dots indicate the momentum values $\sqrt{mE_b/3}$ where $\phi_d$ needs to be evaluated. These results are obtained from a Lennard-Jones potential with 15 $s$-wave molecular states and a scattering length $a = 1.21$ $r_{\rm{vdW}}$. Here, $r_{\rm{vdW}}=\frac{1}{2}(mC_6 / \hbar^2)^{1/4}$ is the characteristic length scale of the van der Waals interaction. Given $r_{\rm{vdW}}=82.64$ a$_0$ for $^{87}$Rb, this scattering length corresponds to $a = 100.36$ a$_0$.    }\label{fig:th}
\end{figure}

In order to analyze the scaling of $L_3(\vib, L_R)$ with the molecular binding energy $E_b$, we use the relation $\frac{3q_d^2}{4m}-E_b=0$ and ignore all coefficients independent on $q_d$ in Eq. (\ref{K31}) to obtain
\begin{equation} \label{K3d}
L_3(\vib, L_R)\propto (E_b)^{1/2}|\phi_d(\sqrt{mE_b/3})|^2|t_{\rm{h}}(2\sqrt{mE_b/3})|^2.
\end{equation}
 In Fig.$\:$\ref{fig:th}(a) we show
 $t_{\rm{h}}(p)$.
 It oscillates but the amplitude varies only slowly with the two-body momentum $p$. Of course, this only holds until the bottom of the interaction potential (corresponding to the most deeply-bound states) is reached, as the bottom leads to a momentum cut-off. 
Figure \ref{fig:th}(b) shows $\phi_d(p)$ which is discussed in the main text.

\begin{thebibliography}{38}
\bibitem [{Las()}]{LastLobe}%
  \BibitemOpen
  \href@noop {} {\bibinfo {title} {Technically, ${N}$ depends on ${E}_b$.
  {T}his dependence, however, is so weak that its effect on the scaling is
  negligible.}}\BibitemShut {Stop}%
\end{thebibliography}
%

\end{document}